\pdfoutput=1 
\documentclass[%
reprint,
superscriptaddress,
 amsmath,amssymb,
 aip,
]{revtex4-1}

\usepackage{graphicx}
\usepackage{dcolumn}
\usepackage{bm}

\usepackage[labelfont=bf,labelsep=period,justification=raggedright]{caption}
\usepackage{setspace}

\begin{document}


\title{How structure sculpts function: unveiling the contribution of anatomical connectivity to the brain's spontaneous correlation structure} 


\author{R. G. Bettinardi}	\email{ruggero.bettinardi@upf.edu}
\affiliation{Center for Brain and Cognition, Universitat Pompeu Fabra, Barcelona, Spain.}
\affiliation{Department of Information and Communication Technologies, Universitat Pompeu Fabra, Barcelona, Spain.} 

\author{G. Deco}
\affiliation{Center for Brain and Cognition, Universitat Pompeu Fabra, Barcelona, Spain.}
\affiliation{Department of Information and Communication Technologies, Universitat Pompeu Fabra, Barcelona, Spain.} 
\affiliation{Instituci\'o Catalana de la Recerca i Estudis Avan\c cats, Universitat Pompeu Fabra, Barcelona, 	Spain.}

\author{V. M. Karlaftis}
\affiliation{Department of Psychology, University of Cambridge, United Kingdom.}

\author{T. J. Van Hartevelt}
\affiliation{Department of Psychiatry, University of Oxford, United Kingdom}
\affiliation{Center for Music in the Brain, Aarhus University, Aarhus, 8000 Aarhus C, Denmark}

\author{H. M. Fernandes}
\affiliation{Department of Psychiatry, University of Oxford, United Kingdom}
\affiliation{Center for Music in the Brain, Aarhus University, Aarhus, 8000 Aarhus C, Denmark}

\author{Z. Kourtzi}
\affiliation{Department of Psychology, University of Cambridge, United Kingdom.}

\author{M. L. Kringelbach}
\affiliation{Department of Psychiatry, University of Oxford, United Kingdom}
\affiliation{Center for Music in the Brain, Aarhus University, Aarhus, 8000 Aarhus C, Denmark}

\author{G. Zamora-L\'opez}  \email{gorka@Zamora-Lopez.xyz}
\affiliation{Center for Brain and Cognition, Universitat Pompeu Fabra, Barcelona, Spain.}
\affiliation{Department of Information and Communication Technologies, Universitat Pompeu Fabra, Barcelona, Spain.} 


\begin{abstract}

Intrinsic brain activity is characterized by highly structured co-activations between different regions, whose origin is still under debate. In this paper, we address the question whether it is possible to unveil how the underlying anatomical connectivity shape the brain's spontaneous correlation structure. We start from the assumption that in order for two nodes to exhibit large covariation, they must be exposed to similar input patterns from the entire network. We then acknowledge that information rarely spreads only along an unique route, but rather travels along all possible paths. In real networks the strength of local perturbations tends to decay as they propagate away from the sources, leading to a progressive attenuation of the original information content and, thus, of their influence. We use these notions to derive a novel analytical measure, $\mathcal{T}$ , which quantifies the similarity of the whole-network input patterns arriving at any two nodes only due to the underlying topology, in what is a generalization of the matching index. We show that this measure of topological similarity can indeed be used to predict the contribution of network topology to the expected correlation structure, thus unveiling the mechanism behind the tight but elusive relationship between structure and function in complex networks. Finally, we use this measure to investigate brain connectivity, showing that information about the topology defined by the complex fabric of brain axonal pathways specifies to a large extent the time-average functional connectivity observed at rest.\\ 
\end{abstract}

\maketitle

\section{Introduction \label{sec:Intro}}

In the last two decades a large body of research has demonstrated that spontaneous brain activity forms structured patterns of consistent co-activations across different subsets of brain regions
~\cite{biswal_functional_1995, arieli_dynamics_1996, gusnard_searching_2001, mazoyer_cortical_2001, shulman_energetic_2004, fox_human_2005, beckmann_investigations_2005, fransson_how_2006, smith_correspondence_2009, biswal_toward_2010}. Contrary to what was somehow implicitly assumed, intrinsic brain activity cannot be considered as a simple sum of rather unpredictable and noisy fluctuations. Spontaneous collective dynamics of different brain regions, measured either with EEG, fMRI or MEG, can be clustered into highly organized and reproducible spatial patterns, referred to as resting-state networks (RSNs), that strikingly resemble those activations observed during the performance of different tasks~\cite{laufs_electroencephalographic_2003, beckmann_investigations_2005, brookes2011investigating}.
Furthermore, growing evidences suggest that this spontaneous large-scale structure is characterized by a marked temporal organization, mirrored by recurrent alternations of subnetworks that concur in generating a rich dynamical repertoire at different time scales~\cite{ponce-alvarez_resting-state_2015, keilholz2016multiscale}. Although the origin and the purpose of spontaneous brain dynamics are still under debate, there is wide agreement that mental states and brain malfunction alter the patterns of spontaneous activity, e.g. the dynamical repertoire of the brain at rest tends to decrease during sleep~\cite{Deco_CortexAsleep_2013} and under anaesthesia~\cite{Hudetz_SpinGlass_2014, Hudetz_Unconsciousness_2014, bettinardi_gradual_2015}.

From its discovery, great efforts have been invested in trying to reproduce resting-state brain activity through the use of computational models, in order to obtain a mechanistic explanation of this intriguing but elusive phenomenon. Early models based on the structural connectomes of cats and macaques extensively explored the emerging patterns of correlations in those networks at different spatial and temporal scales~\cite{Zhou_HierarchicalCat_2006, honey_network_2007, Gomez_FromModular_2010}. With the arrival of structural human connectomes obtained through tractography, computational models could finally attempt to fit the empirical correlation structure observed via resting-state fMRI~\cite{deco_key_2009, deco_emerging_2011, deco_resting_2013, deco_great_2014, messe_relating_2014}. Despite these attempts, we still lack of a precise understanding of the relationship between the shape of the brain's connectome and the emergent patterns of correlations observed during rest. One of the major reasons is that the collective dynamics of a network do not only depend on the shape of the underlying connectivity, but also on the model chosen to simulate the local dynamics of the cortical regions~\cite{schmidt_simulation_2010, messe_relating_2014}.  

In this paper we aim at unveiling what is the precise contribution of the anatomical connectivity on the correlation structure observed during rest. To this aim, we first need to derive a theoretical estimate of the expected correlation between nodes due to the network's topology alone; therefore we will try to avoid, as much as possible, the contribution of other factors. To do so, in Section~\ref{sec:sculpts} we introduce a novel graph theoretical quantity measuring the topological similarity of the entire input profiles that two nodes receive from the whole network. This measure, that we named \emph{topological similarity}, is a generalization of the concept of matching index that explicitly accounts for the fact that in networks information travels along all possible paths, not only along the shortest ones, and that information content tends to decay as it moves away from its source~\cite{latora_efficient_2001, huberman_information_2004, ashton_effect_2005, trusina_communication_2005}. This measure is based on the concept of network's \emph{communicability} introduced by~\citet{estrada_communicability_2008}, a function that quantifies the strength of the influence that one node exerts over another through all paths of any length assuming an exponential decay of influence with path length. We then demonstrate that the topological similarity function, altough based on pure topological information about the underlying path structure of the network, can indeed be used to approximate the expected correlation structure of networks of time-varying coupled units.

In Section~\ref{sec:primitives} we systematically investigate the contribution of three \emph{topological primitives}: the weight of the links, the length of the path and the presence of redundant alternative paths through which information can travel. Topological primitives are fundamental features underlying the network architecture which determine how influence spreads, thus sculpting the similarity of the input profiles of nodes. To do so, we study simple graphs in which these primitive features can be manipulated. We show how these three features are in fact of primary importance in modulating the strength of the influence and the relative topologies into which different nodes are embedded. 

Finally, in Section~\ref{sec:human} we investigate the spontaneous correlation structure of the human brain, demonstrating that taking into account the similarity of the whole-network influences shaped by the underlying anatomical topology it is possible to understand and predict the large-scale functional organization observed during rest.

\section{How topology sculpts the correlation structure of networks}	\label{sec:sculpts}  	

In order to properly address the contribution of network's topology on the emerging spontaneous correlation structure, we first need to acknowledge that the collective behaviour of a set of coupled dynamical units depends on three principal ingredients: ($i$) the structure of the network, ($ii$) the local dynamics of the nodes and ($iii$) the coupling function determining how information is passed from one node to another. In fact, for a fixed network, changing the local dynamical model of the nodes and the coupling function usually leads to different collective dynamics~\cite{Huang_GenericBehaviour_2009, schmidt_simulation_2010, messe_relating_2014}. Therefore, in order to estimate the contribution of structure alone we need to set apart the role of the other two factors. 

Typically, the activity of two nodes exhibits statistical dependence either if they are connected by means of a direct link, or if the aggregate inputs they receive from the entire network are similar, independently of whether there is a link between the two or not. Because information in a network rarely travels exclusively along the shortest paths~\cite{borgatti_centrality_2005, colizza_detecting_2006} but instead diffuses along the whole network, we realise that the total influence of one node over another mainly depends on three topological features: ($i$) the strength of the coupling between them, usually represented by the weights of the links, ($ii$) the graph distance between the two nodes and ($iii$) the propagation of the influence along multiple alternative paths~\cite{goni_resting-brain_2014, Avena_PathEnsembles_2016}. We refer to these three topological features as the \emph{topological primitives} because, in combination, they encode the shape of the network's path structure. As such the analysis of the building blocks underlying larger complex networks is a necessary step to understand the contribution of network structure to the emergent activity.

In general, the influence of a direct link is greater than the influence exerted over longer paths, as the latter is mediated through third nodes. In fact, in real systems the ``power'' of the signals or their information content naturally decays along the path~\cite{latora_efficient_2001, huberman_information_2004, ashton_effect_2005, trusina_communication_2005}, unless there exists an active mechanism which amplifies the incoming signal at the cost of energy. Following the same rationale, it is unlikely that influence or information from one node to another propagates only along a single, selected path, unless there are specific gating mechanisms controlling the detailed routing of information over all existing paths. The total number of paths (of all lengths) between any two nodes in a network is in fact infinite. It is well known that the total number of paths $m^l_{ij}$ of length $l$ between nodes $i$ and $j$ in a graph grows with $l$. This number is given exactly by the $l^{th}$ power of the adjacency matrix $A$, $m_{ij} = (A^l)_{ij}$~\cite{harary_spectral_1979, bang-jensen_digraphs_2008}. Thus, the total number of paths leaving from node $i$  and arriving at node $j$ is given by the sum: 

\begin{equation} 	\label{eq:totalpaths}
\sum_{l=0}^\infty (A^l)_{ij} = \mathbf{1} + A_{ij} + (A^2)_{ij} + (A^3)_{ij} + \ldots
\end{equation}

This number typically diverges and thus, for the dynamics within a network to remain bounded, the \emph{amount of influence} needs to decay faster with the length than the growth in the number of paths. Mathematically, the problem consists in finding a set of coefficients $\{ k_l \}$ for which the series $\sum_{l=0}^\infty k_l \, A^l$ converges for any adjacency matrix, $A$. While the solution to this problem is not unique,~\citet{estrada_communicability_2008} proposed an exponential decay of the influence with path length and introduced the \emph{communicability} measure $\mathcal{C}$. The communicability function thus corresponds to the matrix exponential of $A$, which can be expanded into a series of powers with coefficients  $k_l = 1 / \,l!$:

\begin{equation} 	\label{eq:Communicability}
 \mathcal{C} \equiv e^A = \sum_{l=0}^\infty \frac{A^l}{l!} = \mathbf{1} + A + \frac{A^2}{2!} + \frac{A^3}{3!} + \dots
\end{equation}

From a physical perspective, the communicability is analogous to the Green's function of the network~\cite{estrada_communicability_2008, estrada_physics_2012} and expresses how local perturbations propagate along the system. Perturbation of a given node's dynamics in fact first propagates to its direct neighbours, affecting their activity; the activity of the neighbours will in turn propagate to their neighbours (as well as back to the node that was perturbed in the first place), and interact with their intrinsic local dynamics. This simple propagation mechanism implies that the effect of a local perturbation could be perceived also by distant nodes but attenuated and modulated by the dynamics of each node along the route. Due to its correspondence to the Green's function, the communicability can be tuned using a constant global coupling parameter  that uniformly scales the weights of all links in $A$~\cite{estrada_physics_2012, zamora_functional_2016}, allowing to search through the emerging collective dynamics over multiple scales. The generalised, tunable, communicability is then:

\begin{equation} 	\label{eq:Communicability_g}
 \mathcal{C} = e^{gA} = \sum_{l=0}^\infty \frac{g^lA^l}{l!} = \mathbf{1} + gA + \frac{g^2 A^2}{2!} + \frac{g^3 A^3}{3!} + \dots 
\end{equation}

When $g$ is weak, perturbations quickly decay, producing local correlations only around the node's neighbourhood. As $g$ grows perturbations propagate deeper into the network, giving raise to stronger correlations over more distant nodes. \citet{zamora_functional_2016} have shown that considering the communicability as the propagator kernel for the diffusion of Gaussian noise along the network, it reveals an equivalent correlation structure as networks of generic and widely used models, e.g. Kuramoto oscillators and neural masses (see Supplementary Information in Ref.~\onlinecite{zamora_functional_2016}). Under these assumptions (Gaussian white noise sources and exponential decay), in Ref~\onlinecite{zamora_functional_2016} it was shown that the covariance matrix of the system, $\Sigma$, can be analytically estimated as $\Sigma = \mathcal{C} \cdot \mathcal{C}^T$ where $\mathcal{C}^T$ is the transposed communicability. The cross-correlation matrix $R$ of the system is then calculated, as usual, normalising the rows of the covariances $\Sigma_{ij}$ by the autocovariances $\Sigma_{ii}$ as, $R_{ij} = \Sigma_{ij} / \Sigma_{ii}$. Despite the merit of being analytical, this estimate of the network's cross-correlation matrix still relied on assuming very simple dynamical model, the diffusion of Gaussian white noise, and did not fully  disentangle the unique contribution of the topology. In the present work we want to address the question whether it is possible to estimate the most likely correlation structure that a network gives rise, based only on its topological properties. As mentioned before, it is legitimate to assume that the activities of two nodes will exhibit statistical dependence either if they are connected by means of a direct link, or if the aggregate input they receive from the entire network is similar. With this in mind, we realised that the column vectors $\mathbf{c_j}$ of the communicability matrix $\mathcal{C}$ indeed represent the input profile of the influences node $j$ receives from all nodes in the network along all possible paths, including the influence a node has on itself due to recurrent paths. Therefore, it should be possible to predict analytically the magnitude of the expected correlation between any two nodes in a network by comparing the similarity of their input profiles. The resemblance between two input profiles can in principle be calculated using any measure of similarity between multidimensional vectors. Here we choose the cosine similarity because it returns results bounded between $-1$ and $1$, equivalent to the cross-correlation measure we aim at comparing. Consequently, we define the \emph{topological similarity}, $\mathcal{T}_{ij}$, between two nodes as the cosine of the angle between the corresponding columns $\mathbf{c}_i$ and $\mathbf{c}_j$ in the communicability matrix: 

\begin{equation} 	\label{eq:TopolCorr}
 \mathcal{T}_{ij} = \frac{\langle\mathbf{c_{i}} , \mathbf{c_{j}}\rangle }{\|\mathbf{c_{i}}\| \|\mathbf{c_{j}}\|} ,
\end{equation}

being $\langle$ , $\rangle$ the inner product and $\| \|$ the vector norm. The definition of $\mathcal{T}$ depends uniquely on the topological constraints of the network encoded in the adjacency matrix $A$, plus the realistic assumption (embedded in $\mathcal{C}$) that the influence or the information content decays with the length of the path. Despite being a purely topological measure, we recognized that, when the variance of the nodes is the same for all nodes, $\mathcal{T}$ corresponds exactly to the correlation matrix $R$ of the Gaussian diffusion dynamics studied in Ref~\onlinecite{zamora_functional_2016}. Therefore, $\mathcal{T}$ formally closes the cycle for the search of a direct relation between the structure of a network and the expected correlation pattern that this structure will tend to generate. Furthermore, this analytical measure unveils the fundamental mechanism behind the contribution of network structure to the emerging function, which can be understood in terms of the similarity of influences that two nodes receive from the whole network via its complete path structure.    

Finally, we shall notice that $\mathcal{T}$ can be estimated for both directed and undirected graphs, as well as for weighted networks. In the case of undirected graphs $\mathcal{C}$ is symmetric but if the links are directed, then the columns of $\mathcal{C}$ determine the input profiles and the rows represent the profile of output influences of the nodes. It must be noted that despite the measure can be computed for any weighted adjacency matrix, it \emph{does not} always make sense to do so. Because communicability is a measure of influence along the paths, it only has a direct physical meaning when the weights of the links represent the coupling strength between the nodes, the flow capacity of the link or a compatible physical sense. 

Summarising, here we have introduced a topological estimator of a network's correlation structure. Although it ignores any specific dynamics of the nodes, it accounts for the fact that information or influence within a network propagates along all possible alternative paths and that naturally decays for longer paths. The measure actually quantifies the similarity between the input profiles of nodes. In the following section we systematically investigate how three fundamental features of a network (the weights of the links, the path length and the redundancy of paths) influence both the communicability and the topological similarity between nodes that, as shown above, can in fact be used as a proxy of their expected correlation.

\section{Effect of topological primitives} 	\label{sec:primitives}

In this section, we will investigate how three topological primitives we defined before, namely links' weights, graph distance and the presence of multiple alternative paths, sculpts the influence that one node exerts over another, setting aside the role of local nodes' dynamics. To this aim, we will focus on three simple classes of graphs: ($i$) chains, ($ii$) cycles and ($iii$) path-redundant networks. We will evaluate how manipulating critical parameters of these graphs leads to changes in the influence between given pairs of nodes $a$ and $b$ (measured by their communicability, $\mathcal{C}_{ab}$) and in their topological similarity $\mathcal{T}_{ab}$.

The contribution of link weight in the simplest case can be understood when analyzing how changes in the weight modulate the mutual influence of two nodes directly connected by a single link (see Supplemental Figure S1). As expected, incresing link weight is associated with increase in both the influence that one node exert over the other, as well as in their topological similarity.

\begin{figure*}
\centering
\includegraphics[width=1\textwidth]{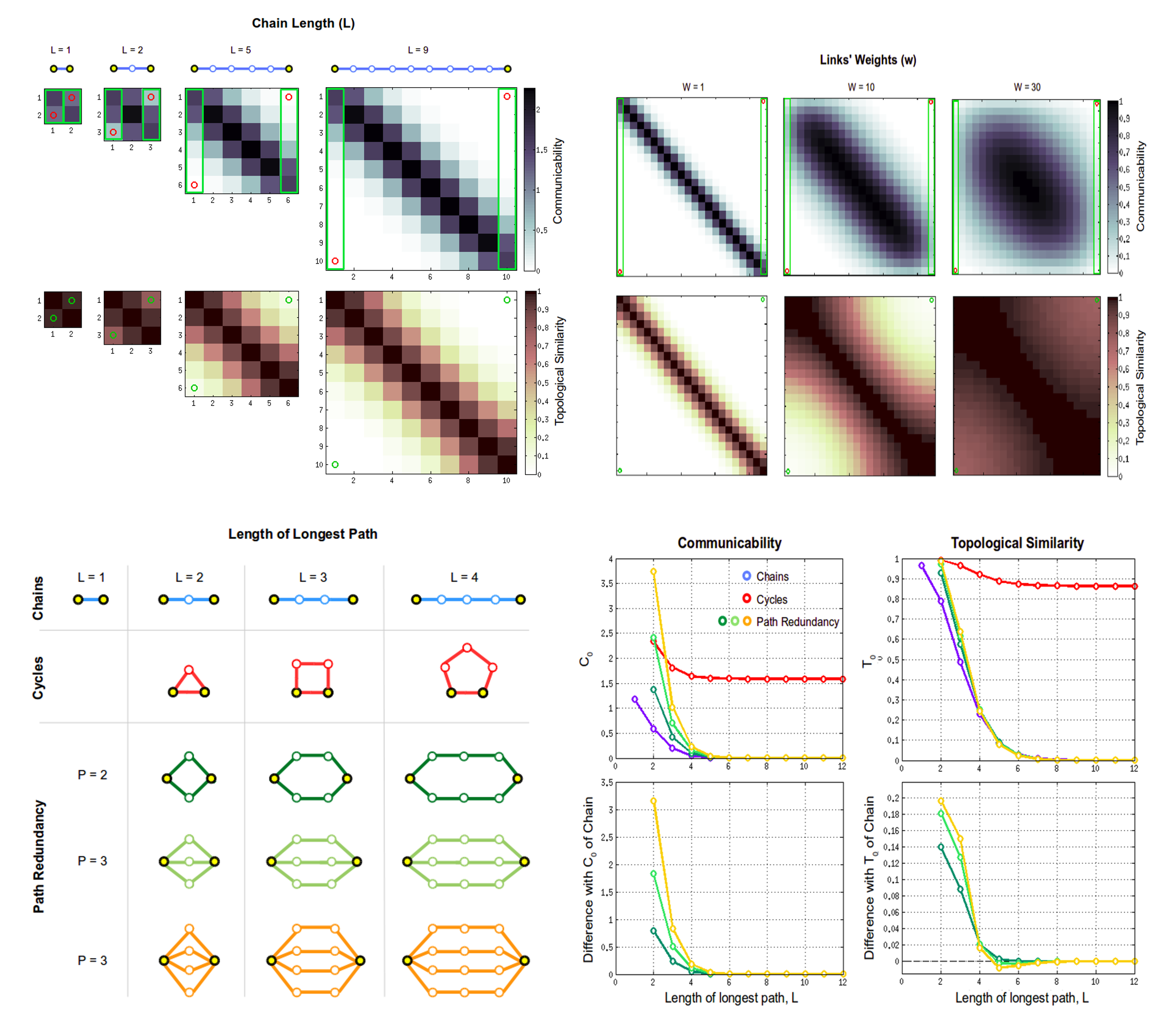}
\caption[Communicability and Topological Similarity in simple network motifs.]{\textbf{Behavior of Communicability and Topological Similarity in simple network motifs.}
(Top-left) Communicability and Topological Similarity matrices of chains of different lengths. In the upper matrices, the red dots indicate the matrix entry corresponding to the communicability between the nodes at the two ends of the chains, whereas the green rectangles marks their whole-network input profiles (column vectors), that are used to calculate the topological similarity of the corresponding nodes (marked with the green dots in the lower matrices.   
(Top-Right) Communicability and Topological Similarity matrices of chains of constant length ($L=$21) for different links' weights, $w$. See top-left caption for the legend of red dots, green rectangles and green dots.
(Bottom-left) Schematic representation of different graphs (chains, cycles and path-redundant architectures) having comparable longest path. The reference nodes for which both the communicability ($\mathcal{C}_{ab}$) and the topological similarity ($\mathcal{T}_{ab}$) where calculated are highlighted in yellow. 
(Bottom-Right) Upper panels: comparison of $\mathcal{C}_{ab}$ and $\mathcal{T}_{ab}$ of the three different graphs having comparable longest path. Line colors correspond to those in the schematic representation in the bottom-left region of the figure (Light blue lines: chains; Red lines: cycles; Dark green lines: two redundant paths; Light green lines: three redundant paths; Orange lines: four redundant paths). 
Lower panels: diffference between $\mathcal{C}_{ab}$ and $\mathcal{T}_{ab}$ of chains and redundant topologies.  All results were obtained for constant links' weights and global coupling $w =g =$ 1.
\label{fig:F1}}
\end{figure*}

A simple manner to obtain a better intuition of the behavior of both communicability and topological similarity is by studying how increasing the graph distance between two reference nodes affects both $\mathcal{C}_{ab}$ and $\mathcal{T}_{ab}$ in simple chain topologies. See top-left panels of Figure~\ref{fig:F1}. 

In general, it is possible to see how increasing the length of the path separating the two nodes correspond to a decrease in both their communicability and topological similarity, behavior directly determined by the decay formalized in the definition of the communicability. From the example, it is indeed evident how, for increasing lenghts of the chain, the input profiles of the two nodes at the ends of the chain (highlighted with green rectangles in the communicability matrices, top-left panels of Figure~\ref{fig:F1}) become more and more antithetic, reflecting opposed whole-network influences. This behavior is captured by the corresponding decrease in the nodes topological similarity. The effect of uniformly varying the weights of all links in a chain is mainly quantitative: in fact, augmenting the weights in chains of fixed length increases the strength of the influence of each node over all other nodes in a way that is inversely proportional to the distance separating them, top-right panel of Figure~\ref{fig:F1}. See also Supplemental Figure S2 to appreciate the interaction of chain length and links' weight on $\mathcal{C}$ and $\mathcal{T}$ between the nodes at the two ends of a chain).

Due to the definition of communicability, the length of the path separating any two nodes as well as the weights of the links are the most important parameters defining the strength of their mutual influence and therefore their topological similarity as well. As such, chain topologies can be thought as a baseline to compare how more complex motifs such as cycles and path-redundant topologies, which indeed incorporate chain topologies, modulate both the resulting communicability and the topological similarity between given nodes.

With this in mind, we compared $\mathcal{C}_{ab}$ and $\mathcal{T}_{ab}$ of the three model graphs (chains, cycles and path-redundant motifs) having corresponding graph diameter, \emph{i.e.} having same longest path $L$. Bottom-left panel of Figure~\ref{fig:F1} provides a schematic representation of different motifs having equivalent longest paths. For the case of chains and path-redundant topologies, we computed $\mathcal{C}_{ab}$ and $\mathcal{T}_{ab}$ for those two nodes that were more distant, in other words, those at the two extremes of (each) path, whereas for cyclic topologies, we always selected two adjacent nodes: this choice allowed us to clearly disentangle the contribution that the direct link has on both the commmunicability and on the topological similarity, above and beyond the modulation produced by chains of different length.

As expected, in chains both $\mathcal{C}_{ab}$ and $\mathcal{T}_{ab}$ decay as the distance between the extremal nodes increases, see bottom-right panels of Figure~\ref{fig:F1}, blue lines. On the other hand the effect of the direct link is well illustrated for the case of cyclic architectures (bottom-right panels in Figure~\ref{fig:F1}, red lines). In fact, the presence of a direct link importantly enhances and poses a lower bound for both $\mathcal{C}_{ab}$ and $\mathcal{T}_{ab}$ , that however exhibit a chain-like decay as the indirect path between them increases. 

This approach makes the difference between chains and cyclic primitives straightforward; however, in order to being able to properly understand how increasing path redundancy affects the communicability and hence topological similarity, it can be of great help to directly quantify the \emph{difference} between $\mathcal{C}_{ab}$ and $\mathcal{T}_{ab}$ calculated from chains and from redundant topologies for comparable path length. The two lowest-right panels in Figure~\ref{fig:F1} illustrate three examples of these differences, namely the cases with two (dark green lines), three (light green lines) and four (orange lines) redundant paths. From this analysis, we appreciate that increasing the number of alternative paths does increase both the total influence and the topological similarity between the two reference nodes at both ends of the paths, but that the magnitude of this increase decays with the length of the path, vanishing for paths longer than 5 links.

Studying these simple networks (single link, chain, cycles and redundant paths) gives the opportunity to understand the central relevance of the three topological primitives in shaping how the influences of nodes unfould through the graph. This information in turn determines how strong will be the expected correlation between any pair of nodes, approximated by their topological similarity. These simplified network models can thus be used to summarize some of the simplest forms of interactions sustained by these topological features.

\section{Understanding the brain's spontaneous correlation structure} 	\label{sec:human}

In the previous section we have analyzed how simple network architectures could affect both communicability and topological similarity between a given pair of nodes. However, real networks are made of interwined assemblies of those topological motifs that form intricate architectures and, together with the particular dynamical properties characterizing the system at hand, determine the emergence of complex patterns of interactions. In this section, we will try to find out) how much of the complex pattern of spontaneous correlations empirically observed in the resting brain can be explained just by the topology of the underlying anatomical structure. 

The main assumption we make is the following: if two brain regions receive similar influences from the entire network, then the probability that they will exhibit consistent co-activations is high; on the other hand, if the inputs they receive from the whole network are very different, it is more likely that their covariance is weak. In Sections~\ref{sec:sculpts} and~\ref{sec:primitives}, we showed how the the structure of a network can be used to estimate the strength of the influence that one node exerts over another trough the communicability $\mathcal{C}$, and that $\mathcal{C}$ can be further used to quantify the topological similarity, $\mathcal{T}$, of the input profiles between any pair of nodes. We then demonstrated that $\mathcal{T}$ can in fact be interpreted as the expected correlation between the nodes in the network. As such, we will use $\mathcal{T}$ to estimate the unique contribution of topology to the empirical time-average correlation structure of the brain's spontaneous activity measured using resting-state fMRI.
 
To this aim, we will compute the topological similarity matrix from the group-average structural connectivity matrix (SC), and use it to estimate the correlation structure mirrored by the group-average empirical functional connectivity matrix (FC) obtained from resting-state fMRI. See Section~\ref{sec:MM} for details about the empirical SC and FC matrices. The SC matrix stores the information about axonal pathways reconstructed using whole-brain diffusion tensor imaging and tractography, thus defines the whole-brain anatomical wiring diagram of the brain. It should be noted that, despite their reproducibility, current methods used to reconstruct fiber bundles from diffusion imaging are characterized by intrinsic limitations constraining their accuracy: in fact, it is well-known that these reconstruction algorithms tend to favor the shortest, straightest and simplest path between any two reference voxels~\citep{jones_challenges_2010}, which in turn impair their ability to accurately detect crossing fibers and long inter-hemispheric axons~\citep{thomas_anatomical_2014, jbabdi_measuring_2015}. With this in mind, we focused only on intra-hemipsheric structural and functional connectivity, in order to avoid the confounding effects of accumulating errors due to the above mentioned limitations in sampling inter-hemispheric pathways.

\begin{figure}
\centering
\includegraphics[width=1\columnwidth]{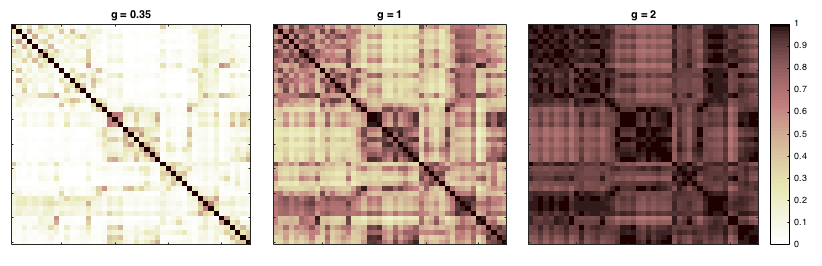}
\caption[Effect of the global coupling]{\textbf{Effect of the global coupling.}
The figure illustrate the effect of the global coupling, $g$, onto the resulting topological similarity matrix, $\mathcal{T}$. The three matrices have been obtained from the same empirical structural connectivity matrix, SC. 
\label{fig:F3}}
\end{figure}

As mentioned in Section~\ref{sec:sculpts}, the communicability can be scaled by a factor $g$ controlling the emerging collective dynamics, thus leading to the possibility of obtaining as many topological similarity matrices as the values of $g$ used to scale $\mathcal{C}$ (see Figure~\ref{fig:F3} for three instances of $\mathcal{T}$ obtained for increasing values of $g$). 

\begin{figure*}
\centering
\includegraphics[width=1\textwidth]{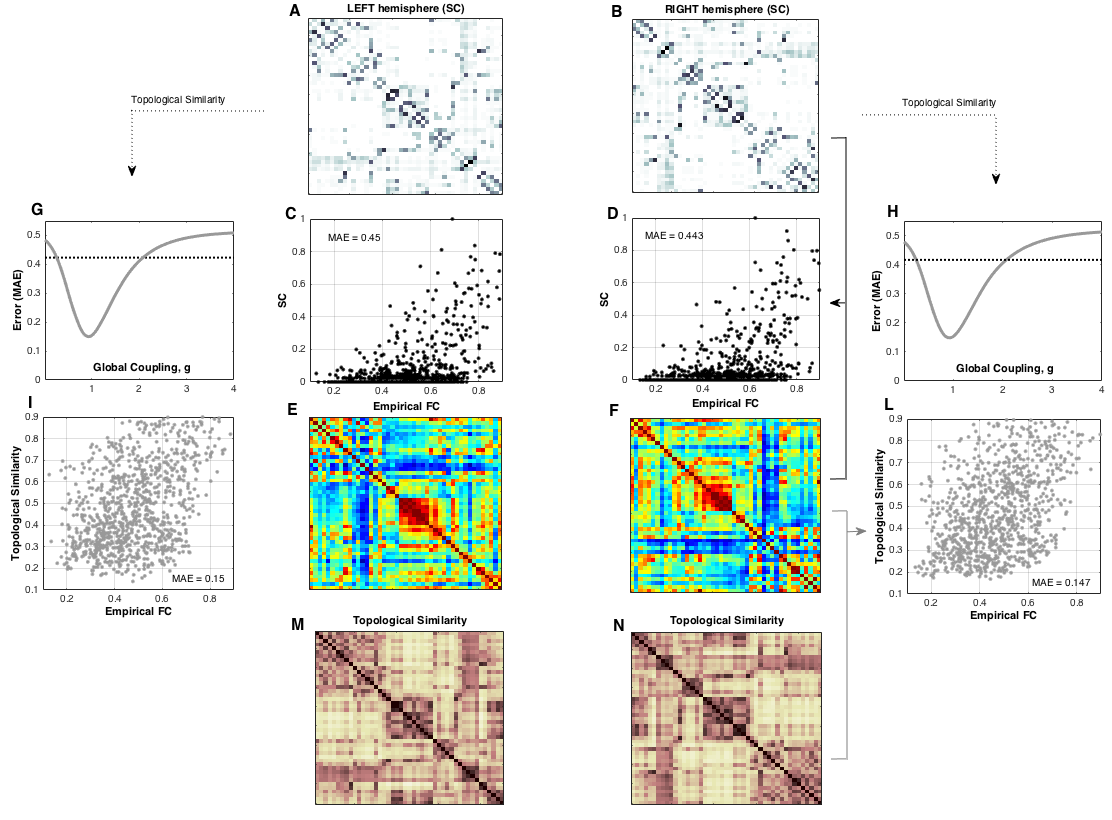}
\caption[Approximate brain's spontaneous correlation structure]{\textbf{Using topology to approximate the brain's spontaneous correlation structure.}
The figure shows results obtained separately for the left and right hemisphere. 
(A,B) Intra-hemispheric structural connectivity matrices (SC).  
(C,D) Scatterplot depicting the relationship between SC and the intra-hemispheric empirical correlation structure, summarized also in panels E and F.
(E,F) Empirical functional connectivity matrices (FC) for the left and the right hemisphere.
(G,H) Mean absolute error (MAE) between the empirical FC and the topological similarity $\mathcal{T}$ computed for different value of the global coupling parameter, $g$. For both hemispheres, minimal MAE ($\approx$0.15) corresponded to $g\approx$1. The dotted lines correspond to the mean absolute error between the SC matrix and the empirical FC, which $\approx$0.45. 
(I,L) Scatterplots of the empirical FC and the best-fitting topological similarity for the two hemispheres.
(M,N) Best-fitting Topological similarity matrices of the two hemispheres. 
\label{fig:EmpFit_T}}
\end{figure*}

For each hemisphere, we thus searched for the topological similarity matrix $\mathcal{T}$ that best explained the observed intra-hemispheric correlation structure, by optimizing the communicability matrix according to a global scaling factor, $g$ (Panel G and H in Figure~\ref{fig:EmpFit_T}). 

As a mesure of model fitting, we used the mean absolute error (MAE), an outlier-robust alternative of the mean squared error (MSE), a classic statistic to quantify the goodness of an estimator. From the scatter plots (Panels I,L of Figure~\ref{fig:EmpFit_T}) and the best-fitting $\mathcal{T}$ matrices (Panels M,N of Figure~\ref{fig:EmpFit_T}), it is possible to appreciate how properly accounting for the overall input pattern sustained by a given topology can indeed reduce the error in estimating the emergent correlations from the raw structural connectivity alone; in fact, in the example we pass from a $E$(SC,FC)$\approx 0.42$ to $E$($\mathcal{T}$,FC)$\approx 0.15$).\\

These results demonstrate that knowledge of the topology of whole-network input patterns of different brain regions, sustained by direct and indirect routes of multiple interweaved axonal bundles, can very much predict the time-average correlation structure observed from spontaneous BOLD fluctuations, above and beyond the information about direct anatomical connections stored in the SC matrices. 

In the following, we will investigate how explicitly introducing local nodes' dynamics could affect the prediction of the average correlations. To this aim, we will make use of the Hopf normal model, that have successfully used to predict mesoscopic brain activity~\cite{freyer_biophysical_2011,deco_metastability_2016}, and evaluate their ability to explain brain functional connectivity. By this comparison, we will be able to gain some insight about the contribution of adding local dynamics in determining the observed correlation structure, as well as to have a better understanding of the role of the underlying topology, shared either by the two models. 

\subsection{Introducing local dynamics} \label{sec:hopf}

\begin{figure}
\centering
\includegraphics[width=1\columnwidth]{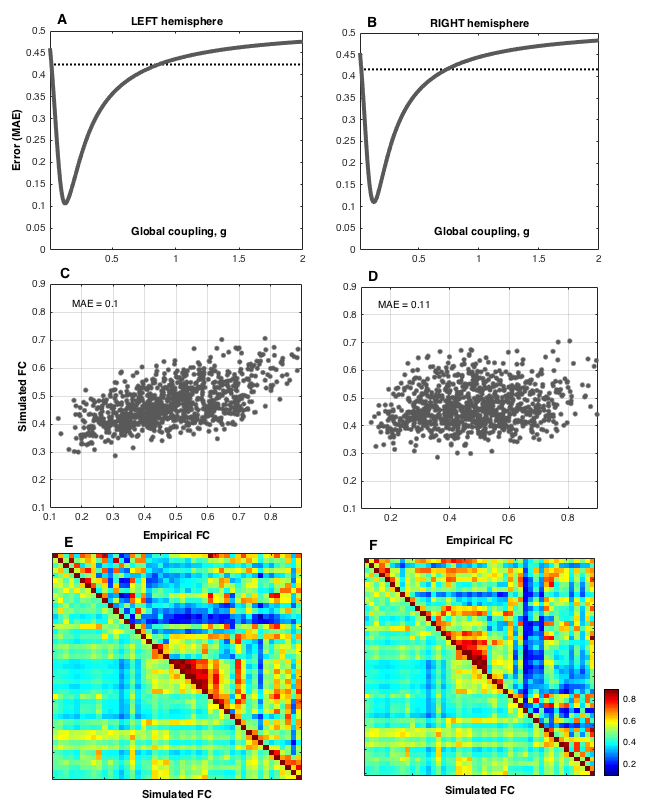}
\caption[Numerical simulations]{\textbf{Numerical simulations.}
The figure illustrate results for the two hemispheres obtained from numerical simulations using the Hopf normal model. 
(A,B) The model returned best results (MAE$\approx$0.1) for values of the global coupling, $g \approx 0.12$ for both hemispheres. As for data shown in Section~\ref{sec:human}, model fitting was performed using the mean absolute error (MAE). MAE between empirical structural (SC) and functional connectivity (FC) matrices is indicated by the dotted lines ($\approx0.42$ in both hemispheres).
(C,D) scatter plots of the empirical functional connectivity (FC) versus the simulated one obtained at the best-fitting global coupling. 
(E,F) In both panels, the upper triangles store the empirical FC values, whereas the lower triangles the corresponding ones obtained from simulations at the best fitting value of $g$.  
\label{fig:hopf}}
\end{figure}

We will now explicitly introduce local dynamics to simulate large-scale brain activity using the connectional architecture defined by the empirical SC matrix. By doing this, we will be able to estimate, through numerical simulations, the correlation between different brain regions, and then compared these results with the average empirical correlation matrix (Empirical FC) obtained from resting-state fMRI. Numerical simulations will be achieved through the use of the Hopf normal model, a dynamical model able to reconcile noise-based approaches with models based on oscillators. This formalism is based on the normal form of a Hopf bifurcation~\cite{freyer_biophysical_2011,freyer_canonical_2012}, a type of bifurcation that occurs when a system characterized by a stable fixed point loses its stability by exhibiting oscillations. As such, this model allows transitions between asynchronous noise activity and oscillations, thus making it a good candidate to reproduce empirical data as observed either with EEG, MEG or fMRI~\cite{freyer_biophysical_2011,freyer_canonical_2012,deco_metastability_2016}. This model rely on the choice of two parameters, namely $g$, the global scaling of the strength of all the links in the network, and $\alpha$, the parameter controlling the dynamical regime of each node. The bifurcation parameter will be set to 0, meaning that all nodes lie at the bifurcation working point, region that has been demonstrated to give good approximate of the empirical resting-state FC~\cite{deco_metastability_2016}, and we will thus optimize for $g$. Results from numerical simulations are depicted in Figure~\ref{fig:hopf}. The introduction of complex local dynamics does indeed increase the capacity of predicting the time-average correlation structure (MAE($\mathcal{T}$)$\approx$ 0.15, whereas MAE(Hopf)$\approx$ 0.11)), even though the magnitude of this increase is relatively small, suggesting that most of the average structure observed in correlated spontaneous BOLD fluctuations are in fact largely determined by the topology of the underlying network's architecture.

\section{Summary and Discussion	\label{sec:summary} } 	

The brain is a complex system and as such its overall dynamics cannot be fully understood without taking into account the rich patterns of interactions into which its components are inherently embedded into. The collective dynamics in a network results from the complex interplay between its underlying structure, the nature of the local dynamics of individual nodes, their specific working point and the manner in which they are coupled~\cite{Huang_GenericBehaviour_2009, schmidt_simulation_2010, messe_relating_2014}. Here, we aimed at understand to what extent does the structure of a network drive the emerging patterns of interactions. Is it possible, knowing the complete wiring diagram of a network, to estimate its most likely correlation structure?

Setting aside the large influence from the local dynamics, we propose that the topological features which play major roles in shaping the interaction between two nodes are: ($i$) the strength of the links, ($ii$) the length of the path between the nodes, and ($iii$) the routing of information along multiple and redundant paths. A strong direct connection between two nodes is usually a reliable indicator of the strength of their functional interaction. In fact, in general direct links lead to more effective communications since the content of information (e.g., the amplitude of the perturbations) tends to decay over longer paths~\cite{latora_efficient_2001}. However, the flow of information from one node to another does not follow a unique path, but rather spreads along several. Therefore, the total influence that one node exerts over another is accumulated over all possible paths, of all lengths~\cite{huberman_information_2004, ashton_effect_2005, trusina_communication_2005}. \\

The goal of this paper is that of answering the question of how the structure of a network contributes to sculpt the expected pattern of correlations that emerge when the network hosts a dynamical process. A direct application is that of understanding the extent to which the structural connectivity of the brain determines the large-scale correlation structure observed during rest. Accordingly, we have developed a graph measure, the \emph{topological similarity} $\mathcal{T}$, which estimates the expected correlation between two nodes based on the similarity of the ``influences'' both receive from the whole network. If two nodes receive the same sets of inputs, then they should be strongly correlated. On the contrary, if they share no common inputs, their correlation tends to be weaker. Hence, this measure can be considered as a generalization of the \emph{matching index}. In the analysis of graphs the matching index is typically used to quantify the number of common neighbours shared by two nodes. However, the matching index only accounts for the direct neighbours and ignores the rest of the complex fabric of interactions sustained by the entire network. To perform a complete comparison of the inputs accumulated over all paths, while accounting for the natural decay of signal power or information content in any real system, we considered the columns of the communicabiilty measure~\cite{estrada_communicability_2008} as they directly correspond to the whole-network input profiles of the nodes. Although the actual decay rate of the signals may differ across real systems, our choice of the communicability guarantees (due to its exponential decay) that the accumulation of perturbations over all possible paths, Eq.~(\ref{eq:totalpaths}), converges for all adjacency matrices $A$. However, any other formalism implementing a decay of the influence between nodes as a function of their graph distance can be used to compute the topological simmilarity $\mathcal{T}$, with the constrain that such formalism must be based on a set of coefficients assuring series convergence for any adjacency matrix.  

Recently, \citet{zamora_functional_2016} found that considering the communicability as the propagator kernel for the diffusion of Gaussian noise along the network allows to analytically estimate the time-average cross-correlation matrix $R$ of the system. We realise that both approaches, the one starting from a dynamical system describing the propagation of perturbations\cite{zamora_functional_2016}, and the other based uniquely on topological constrains and the assumption of exponential decay of influences, are indeed equivalent. In fact, we find that $\mathcal{T} \equiv R$ when the variance $\xi_i$ of the Gaussian noise is the same for all nodes in the Gaussian diffusion system. Therefore we can conclude, with a high confidence, that $\mathcal{T}$ represents the most likely expected correlation structure of a network only due to its underlying topology. 

As argued before, the other crucial ingredients for the collective dynamics on a network are the local dynamics of the nodes and the coupling functions, both of which can be either linear or nonlinear. The topological similarity $\mathcal{T}$ thus captures the \emph{tendency} of the nodes of the network to correlate or to synchronise with each other. The contribution of the local dynamics and of the coupling function is to modulate this original background tendency set by the connection topology and either enhance or disrupt the underlying patterns of correlation.

In order to gain understanding on how each of the three topological features, namely, link weight, path length and path redundancy, precisely affect how influence propagates through the network, we have first investigated the behaviour of $\mathcal{C}$ and $\mathcal{T}$ between selected nodes in simple networks: chains and cycles of varying length, and path-redundant graphs. Not surprisingly, we have found that the strength of a direct link between two nodes is a major contributor to the intensity of their expected correlation. However, the presence of common inputs or redundant paths between them enhances the intensity of their interaction beyond the baseline determined by the strength of the link. If there is no direct link between two nodes, common inputs and redundant paths can trigger strong correlations between them. However, this influence tends to decay with the length of the paths. Although the precise decay rate of the influence depends on the characteristics of the real system, it is worth noting that real networks with diameter larger than five are rather uncommon. Specially in brain and neural networks which are dense. From a more general perspective, we shall notice that large efforts have been devoted in the literature to investigate which of the common network properties, e.g., degree distribution, clustering coefficient, communities or motifs, determine more prominently the collective dynamics on a network, particularly its capacity to synchronise~\cite{Motter_Enhancing_2005, Bernardo_Synch_2005, Atay_Spectra_2006,  Zhou_HierarchicalSynch_2006,  Wu_SynchroClust_2006, Arenas_Review_2008}. After our observations we can conclude that what truly matters for the collective dynamics is the path structure of the network which determines how far do perturbations reach and how their effect accumulates over redundant and recurrent paths. Thus, we foresee that, to precisely understand how the typical network properties determine the collective dynamics, future work needs to identify how those properties alter the underlying network's path structure.

\subsection*{Limitations and outlook}

The results we have here presented come with some limitations which shall be underlined. ($A$) The topological similarity $\mathcal{T}$ we have introduced represents an estimation of the time-averaged cross-correlation. Thus, it only estimates the spatial correlations between brain regions and does not capture temporal correlations. ($B$) Communicability measure assumes an exponential decay of the influence with pathlength. This is not the only choice possible and other sets of coefficients $k_l$ exists which will lead the weighted version of the sum in Eq.~(\ref{eq:totalpaths}) converge. The precise decay rate of influence, perturbations or information in real systems will vary according to the system's nature. Thus, in some real applications it might be possible to identify the decay rate with pathlength and identify the ``right'' set of coefficients $k_l$. As such, $\mathcal{T}$ is not restricted by the set of coefficients used in the communicability measure proposed by~\cite{estrada_communicability_2008} .($C$) We found here more convenient to quantify topological similarity as the cosine similarity of the input profiles, Eq.~(\ref{eq:TopolCorr}), as we found it more accurate in estimating the similarity between input profiles of small size, as in the case of short chains. However, other measures are possible, e.g., correlation or euclidean distance, without conceptually modifying the meaning of topological similarity. ($D$) for the analysis on the brain's connectivity and spontaneous correlations, we considered both hemispheres as if they were independent. The reason was to avoid biases due to the unreliability of tractography to identify inter-hemispheric fibers. Still, our comparisons are biased to some extent because the SC-based $\mathcal{T}$ and simulations consider both hemispheres as independent while the empirical resting-state measurements reflect the activity of the brain regions, which are, certainly, embedded on the whole network. Finally, we want to highlight the usefulness of both topological similarity $\mathcal{T}$ and the correlation matrix $R$ of the Gaussian diffusion system~\cite{zamora_functional_2016} to explore the functional connectivity of synthetic and empirical networks. Because they estimate the expected correlation analytically, they are very fast to compute. For example, they allow to compare the effects of network perturbations, e.g., node or link lessions, without the need to run computationally expensive simulations. The only difference between $\mathcal{T}$ and $R$ is that the formalism of $R$ allows also to explore the effects of simulated inputs by increasing of decreasing the variance of the Gaussian noise at selected nodes.

\section{Materials and Methods} \label{sec:MM}

To obtain the population average structural and functional connectomes twenty-one healthy volunteers (mean age 21.56 years; standard deviation 1.84 years; all males; all right handed) participated in five (5) resting-state and two (2) DTI scanning sessions. All participants signed informed consent to participate. The study was conducted at the School of Psychology, Birmingham and was approved by the University of Birmingham Ethics Committee.

\subsection*{Data acquisition}

The scanning sessions were conducted at the Birmingham University Imaging Centre using a 3T Philips Achieva MRI scanner with a 32-channel SENSE head coil. T1-weighted anatomical data (175 slices; $1 \times 1 \times 1$ mm$^3$ resolution) were collected during the first session only. DTI data were collected in two sessions (23.3 ± 2.5 days apart). The DTI acquisition consisted of 60 isotropically-distributed diffusion weighted directions (b=1500 smm-2; TR=9.5s; TE=78ms; 75 slices; $2\times2\times2$ mm$^3$ resolution) plus a single volume without diffusion weighting ($b=0$ smm-2, denoted as b0). The DTI sequence was repeated twice during each session, once following the Anterior-to-Posterior phase-encoding direction and once the Posterior-to-Anterior direction, to correct for susceptibility-induced geometric distortions~\cite{andersson_howto_2003}. Resting-state data were collected in five sessions (the first and the last collected in the same scanning session as the DTI data) using whole brain echo-planar imaging (EPI) (TR = 2s; TE = 35ms; 32 slices; $2.5\times2.5\times4$ mm$^3$ resolution). Participants were instructed to have their eyes open and maintain fixation to a white dot presented at the centre of the screen.

\subsection*{Whole-brain DTI tractography}

We processed the DTI data in FSL version 5.0.8 (FMRIB Software Library, http://fsl.fmrib.ox.ac.uk/fsl/fslwiki/) on a Red Hat Linux operating system. We corrected the data for susceptibility distortions, eddy currents and motion artifacts~\cite{andersson_integrated_2016}. We subsequently rotated the gradient directions (bvecs) to correct them for motion rotation~\cite{leemans_bmatrix_2009, jones_challenges_2010, Ersoz_quantitative_2014}. We then used the corrected gradients in the Bayesian Estimation of  Diffusion Parameters Obtained using Sampling Techniques (BedpostX) tool to generate a distribution model in each voxel~\cite{behrens_characterization_2003}. We used the default parameters in BedpostX for the diffusion modelling: 2 fibers per voxel, weight of 1 for the secondary fibers, discard of the first 1000 iterations before sampling.

We parcellated the brain into 116 areas using the Automated Anatomical Labeling (AAL) atlas~\cite{tzourio_AALparcellation_2002}. We followed a 4-step registration procedure to align the AAL atlas from MNI to native space: (a) align the non-weighted diffusion volume (b0) of each session to their midspace and create a midspace-template, (b) align the midspace-template to the anatomical (T1) scan, (c) align the T1 to the MNI template of FSL, and (d) invert and combine all the transformation matrices of the previous steps to obtain the MNI-to-native registration. The results of each step were visually inspected to ensure that the alignment was successful. Step (a) controls for potential bias towards the first session, when the T1 was acquired (similar methodology to Ref~\onlinecite{smith_normalized_2001}. The first two registrations are 6-dof linear transformations (rigid-body) since we aligned images of the same subject, while the third is 12-dof non-linear to warp the participantÕs brain around the MNI template. The final matrix of step (d) was applied to the AAL atlas using nearest-neighbour interpolation to preserve the labels of the areas.

We calculated the number of probabilistic streamlines starting from each AAL area and reaching any other AAL area by feeding the BedpostX model to the Probabilistic Tracking algorithm (ProbtrackX)~\cite{behrens_probDTI_2007}. The parameters we used in ProbtrackX are: 5000 samples per voxel, 2000 steps per sample until conversion, 0.5mm step length, 0.2 curvature threshold, 0.01 volume fraction threshold and loopcheck enabled to prevent streamlines from forming loops. We normalised the number of streamlines by the size of the seed area and thresholded streamlines lower than 1\% of maximum (i.e. setting them to zero). We subsequently computed the undirected structural connectivity matrix by averaging the normalised streamlines from area $i$ to area $j$ and from area $j$ to area $i$. The results for each subject (in each DTI session) were organised into 42 weighted adjacency matrices $A$ of size $116 \times 116$.

\subsubsection*{Population-average structural connectome}

To estimate the population average structural connectivity (SC) we pooled the 42 SC matrices together (2 per subject) and considered only the reduced parcellation into 90 brain areas (45 per hemisphere) by excluding the 26 regions of the cerebellum and the vermis. The 42 SC matrices contained a variable number $L$ of undirected links ranging from $L = 895$ for the sparsest case (density $\rho = 0.22$) to $L = 1279$ for the densest ($\rho = 0.32$). We noticed that the simple average of the matrices into a single SC matrix by averaging the 42 values each link takes along the pool leads to an average connectome with strongly biased network properties. For example, this plain average SC matrix contained $L = 1967$ links, which is almost twice the number of links as in the individual matrices. In order to avoid this problem we have devised a method which automatically removes outlier links before performing the average. For each link $(i,j)$ we have initially a set of 42 weigths $\{w_{ij}^s\}$ where $s = 1, 2, \ldots, 42$. The method searches for outlier weights (data-points falling out of 1.5 times the inter-quartile range) and removes them from the data pool. The search is iteratively repeated until no further outliers are detected and then the population-average SC weight for the link $(i,j)$ is calculated as the average weight of the surviving values. In practice, the method converges very rapidly and it rarely performs more than 2 iterations per link. This method allows to clean the data without having to set an arbitrary hard threshold~\cite{reus_falsepositives_2013} for the minimally accepted prevalence of the link. Full details of the method are currently in preparation and will be presented somewhere else. The resulting population-average SC matrix out of our iterative pruning method contains $L = 1189$ links ($\rho = 0.30$), which lies within the range of connectivity for the individual 42 matrices. For the simulations we treated the left and the right hemispheres independently, as two matrices of $N = 45$ ROIs because tractography is known to largely miss interhemispheric connections.

\subsection*{Resting-state time-courses and functional connectome}

We pre-processed the EPI resting-state data in FSL version 5.0.8 (FMRIB Software Library, http://fsl.fmrib.ox.ac.uk/fsl/fslwiki/) on a Red Hat Linux operating system using MELODIC (Multivariate Exploratory Linear Optimized Decomposition into Independent Components). We corrected the data for motion and slice scan timing, removed the non-brain tissue, applied 5mm FWHM spatial smoothing and removed spike motion artifacts using WaveletDespike~\cite{patel_wavelet_2014}. We subsequently applied high-pass temporal filtering and then extracted the average time-course from each AAL area. To estimate the population-average functional connectivity (FC) matrix we considered again only $N = 90$ brain areas excluding the cerebellum and the vermis. We concatenated the 105 sequences of resting-state signals (21 subjects, 5 sessions per subject) into a single long multivariate time-series and computed the Pearson correlation (z-Fisher corrected) for every pair of signals. The opposite procedure, to compute an FC matrix per session and averaging over the 105 FC matrices leads to almost identical results.

\subsection*{Hopf normal model}

Within this model, the temporal evolution of the activity $z$ of node $j$ is given in the complex domain as:

\begin{eqnarray}
\frac{dz_j}{dt} & = & [\alpha_j + i\omega_j - |z^2| ] + \sigma\eta_j(t) \\
z_j & = & \rho_je^{i\theta_j}} = {x_j + iy_j
\end{eqnarray}

Where $\omega$ is the node's intrinsic frequency of oscillation, $\alpha$ is the local bifurcation parameter (local because the model allows the possibility to assign a different value of $\alpha$ for each node in the network) and $\eta$ is additive Gaussian noise with standard deviation $\sigma$. This system above shows a supercritical bifurcation at $\alpha$ = 0. Specifically, if 
$\alpha_j$ is smaller than 0, then the local dynamic has a stable fixed point at $z_j$ = 0, while for $\alpha_j$ values larger than 0 there exists a stable limit-cycle oscillation of frequency $f = \omega/2\pi$ . Whole-brain dynamics are described by the following coupled equations:

\begin{eqnarray}
\frac{dx_j}{dt} & = & [\alpha_j - x_j^2 - y_j^2]x_j - \omega_jy_j + \nonumber \\ 
		&& g \sum_{i=1}^N C_{ij} (x_i-x_j) + \sigma\eta_{xj}(t) \\
\frac{dy_j}{dt} & = & [\alpha_j - x_j^2 - y_j^2]y_j + \omega_jx_j + \nonumber \\
		&& g \sum_{i=1}^N C_{ij} (y_i-y_j) + \sigma\eta_{yj}(t) 
\end{eqnarray}

Where $C_{ij}$ is the anatomical connectivity between nodes $i$ and $j$, $g$ is the global coupling factor and 
the standard deviation of gaussian noise is $\sigma$ = 0.02. In this model the simulated activity corresponds to the 
BOLD signal of each node. The intrinsic frequency of each node was estimated as the peak frequency in the associated 
narrowband (\emph{i.e.}, 0.04 - 0.07 Hz~\cite{glerean_functional_2012}) of the empirical BOLD signals of each brain region. We simulated, for ech of the two hemispherese (45 ROIs each), 330000 points using Euler's method for integration (dt = 0.001). The connectivity between all the regions of interest was defined using the empirical structural connectivity matrix (SC), and obtained timeseries were then used to compute the simulated correlation matrix (Simulated FC).

\acknowledgements{
We would like to thank Rui Wang and Caroline di Bernardi Luft for their help in collecting the data used in this study. This work was supported by (RGB) the FI-DGR scholarship of the Catalan Government through the Ag{\`e}ncia de Gesti{\'o} d'Ajuts Universitari i de Recerca, under agreement 2013FI-B1-00099, (GZL) the European Union's Horizon 2020 research and innovation programme under grant agreement No. 720270 (HBP SGA1), (GD) the European Research Council Advanced Grant: DYSTRUCTURE (295129) and the Spanish Research Project PSI2013-42091-P, (ZK) European Community's Seventh Framework Programme [FP7/2007-2013] under agreement PITN-GA-2011-290011, (VMK) European Community's Seventh Framework Programme [FP7/2007-2013] under agreement PITN-GA-2012-316746 and (MLK) by the European Research Council Consolidator Grant: CAREGIVING (615539).
}

%


\newpage
\clearpage
\widetext

\setcounter{equation}{0}
\setcounter{figure}{0}
\setcounter{table}{0}
\setcounter{page}{1}
\setcounter{section}{0}
\makeatletter
\renewcommand{\theequation}{S\arabic{equation}}
\renewcommand{\thefigure}{S\arabic{figure}}
\renewcommand{\bibnumfmt}[1]{[S#1]}
\renewcommand{\citenumfont}[1]{S#1}

\linespread{1.5}

\begin{center}
\textbf{\large \emph{Supplementary material for:} \\ ``How structure sculpts function: unveiling the contribution of anatomical connectivity to the brain's spontaneous correlation structure.'' by }

\vspace{0.5cm}
Ruggero G. Bettinardi, Gustavo Deco, Vasilis M. Karlaftis, Timothy J. Van Hartevelt, Henrique M. Fernandes, Zoe Kourtzi, Morten L. Kringelbach and Gorka Zamora-L\'opez

\end{center}

\vspace{0.5cm}

\clearpage


\begin{figure*}
\centering
\includegraphics[width=1\textwidth]{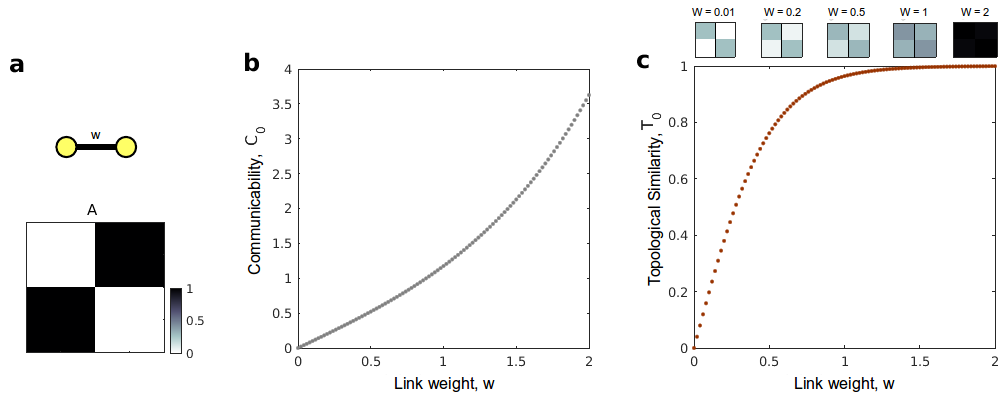}
\caption[Effect of Single Link Weight]{\textbf{Effect of Single Link Weight.}
(A) graph and matrix representation of two nodes connected by a direct link. 
(B) Communicabity between the two nodes, $\mathcal{C}_{ab}$, for diffferent values of link's weight.
(C) Topological Similarity between the two nodes, $\mathcal{T}_{ab}$, for diffferent values of link's weight. On top of the panel are exemplified the communicability matrices obtained for different values of link's weight. Remember that $\mathcal{T}_{ab}$ is computed as the cosine similarity between the two columns of the communicability matrix. 
\label{fig:s1}}
\end{figure*}

\begin{figure*}
\centering
\includegraphics[width=1\textwidth]{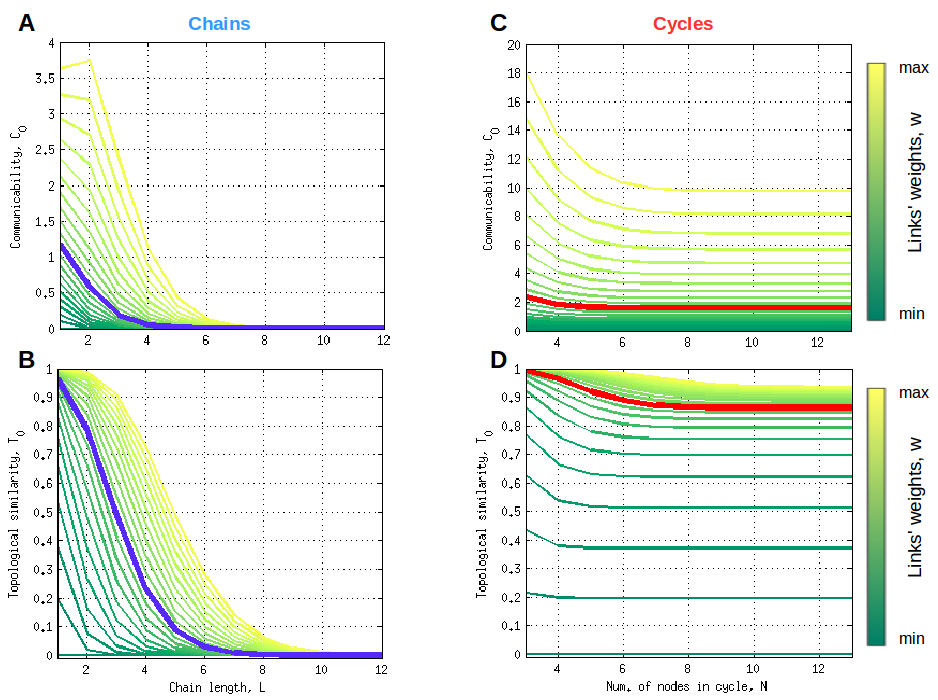}
\caption[Effect of Link Weight]{\textbf{Effect of Link Weight in chain and cyclic topologies.}
(A) Communicability $\mathcal{C}_{ab}$ between the two nodes $a$ and $b$ at the ends of chains of different length for varying links' weights.
(B) Topological similarity $\mathcal{T}_{ab}$ between the two nodes $a$ and $b$ at the ends of chains of different length for varying links' weights. The blue line in Panels A and B correspond to the links' weight value used in chain topologies in the main text of the paper ($w=$1.)
(C) Communicability $\mathcal{C}_{ab}$ between two adjacent nodes $a$ and $b$ for cycles of increasing perimeter ($N$, number of nodes in the cycle) for varying links' weights.
(D) Topological similarity $\mathcal{T}_{ab}$ between two adjacent nodes $a$ and $b$ for cycles of increasing perimeter ($N$, number of nodes in the cycle) for varying links' weights. The red line in Panels C and D correspond to the links' weight value used in cyclic topologies in the main text of the paper ($w=$1.)
\label{fig:s2}}
\end{figure*}



\begin{thebibliography}{64}%
\makeatletter
\providecommand \@ifxundefined [1]{%
 \@ifx{#1\undefined}
}%
\providecommand \@ifnum [1]{%
 \ifnum #1\expandafter \@firstoftwo
 \else \expandafter \@secondoftwo
 \fi
}%
\providecommand \@ifx [1]{%
 \ifx #1\expandafter \@firstoftwo
 \else \expandafter \@secondoftwo
 \fi
}%
\providecommand \natexlab [1]{#1}%
\providecommand \enquote  [1]{``#1''}%
\providecommand \bibnamefont  [1]{#1}%
\providecommand \bibfnamefont [1]{#1}%
\providecommand \citenamefont [1]{#1}%
\providecommand \href@noop [0]{\@secondoftwo}%
\providecommand \href [0]{\begingroup \@sanitize@url \@href}%
\providecommand \@href[1]{\@@startlink{#1}\@@href}%
\providecommand \@@href[1]{\endgroup#1\@@endlink}%
\providecommand \@sanitize@url [0]{\catcode `\\12\catcode `\$12\catcode
  `\&12\catcode `\#12\catcode `\^12\catcode `\_12\catcode `\%12\relax}%
\providecommand \@@startlink[1]{}%
\providecommand \@@endlink[0]{}%
\providecommand \url  [0]{\begingroup\@sanitize@url \@url }%
\providecommand \@url [1]{\endgroup\@href {#1}{\urlprefix }}%
\providecommand \urlprefix  [0]{URL }%
\providecommand \Eprint [0]{\href }%
\providecommand \doibase [0]{http://dx.doi.org/}%
\providecommand \selectlanguage [0]{\@gobble}%
\providecommand \bibinfo  [0]{\@secondoftwo}%
\providecommand \bibfield  [0]{\@secondoftwo}%
\providecommand \translation [1]{[#1]}%
\providecommand \BibitemOpen [0]{}%
\providecommand \bibitemStop [0]{}%
\providecommand \bibitemNoStop [0]{.\EOS\space}%
\providecommand \EOS [0]{\spacefactor3000\relax}%
\providecommand \BibitemShut  [1]{\csname bibitem#1\endcsname}%
\let\auto@bib@innerbib\@empty
\bibitem [{\citenamefont {Biswal}\ \emph {et~al.}(1995)\citenamefont {Biswal},
  \citenamefont {Zerrin~Yetkin}, \citenamefont {Haughton},\ and\ \citenamefont
  {Hyde}}]{biswal_functional_1995}%
  \BibitemOpen
  \bibfield  {author} {\bibinfo {author} {\bibfnamefont {B.}~\bibnamefont
  {Biswal}}, \bibinfo {author} {\bibfnamefont {F.}~\bibnamefont
  {Zerrin~Yetkin}}, \bibinfo {author} {\bibfnamefont {V.~M.}\ \bibnamefont
  {Haughton}}, \ and\ \bibinfo {author} {\bibfnamefont {J.~S.}\ \bibnamefont
  {Hyde}},\ }\bibfield  {title} {\enquote {\bibinfo {title} {Functional
  connectivity in the motor cortex of resting human brain using echo-planar
  {MRI}},}\ }\href@noop {} {\bibfield  {journal} {\bibinfo  {journal} {Magn.
  Reson. Med.}\ }\textbf {\bibinfo {volume} {34}},\ \bibinfo {pages} {537--541}
  (\bibinfo {year} {1995})}\BibitemShut {NoStop}%
\bibitem [{\citenamefont {Arieli}\ \emph {et~al.}(1996)\citenamefont {Arieli},
  \citenamefont {Sterkin}, \citenamefont {Grinvald},\ and\ \citenamefont
  {Aertsen}}]{arieli_dynamics_1996}%
  \BibitemOpen
  \bibfield  {author} {\bibinfo {author} {\bibfnamefont {A.}~\bibnamefont
  {Arieli}}, \bibinfo {author} {\bibfnamefont {A.}~\bibnamefont {Sterkin}},
  \bibinfo {author} {\bibfnamefont {A.}~\bibnamefont {Grinvald}}, \ and\
  \bibinfo {author} {\bibfnamefont {A.}~\bibnamefont {Aertsen}},\ }\bibfield
  {title} {\enquote {\bibinfo {title} {Dynamics of ongoing activity:
  explanation of the large variability in evoked cortical responses},}\
  }\href@noop {} {\bibfield  {journal} {\bibinfo  {journal} {Science}\ }\textbf
  {\bibinfo {volume} {273}},\ \bibinfo {pages} {1868--1871} (\bibinfo {year}
  {1996})}\BibitemShut {NoStop}%
\bibitem [{\citenamefont {Gusnard}\ and\ \citenamefont
  {Raichle}(2001)}]{gusnard_searching_2001}%
  \BibitemOpen
  \bibfield  {author} {\bibinfo {author} {\bibfnamefont {D.~A.}\ \bibnamefont
  {Gusnard}}\ and\ \bibinfo {author} {\bibfnamefont {M.~E.}\ \bibnamefont
  {Raichle}},\ }\bibfield  {title} {\enquote {\bibinfo {title} {Searching for a
  baseline: Functional imaging and the resting human brain},}\ }\href@noop {}
  {\bibfield  {journal} {\bibinfo  {journal} {Nat. Rev. Neurosci.}\ }\textbf
  {\bibinfo {volume} {2}},\ \bibinfo {pages} {685--694} (\bibinfo {year}
  {2001})}\BibitemShut {NoStop}%
\bibitem [{\citenamefont {Mazoyer}\ \emph {et~al.}(2001)\citenamefont
  {Mazoyer}, \citenamefont {Zago}, \citenamefont {Mellet}, \citenamefont
  {Bricogne}, \citenamefont {Etard}, \citenamefont {Houd{\'e}}, \citenamefont
  {Crivello}, \citenamefont {Joliot}, \citenamefont {Petit},\ and\
  \citenamefont {Tzourio-Mazoyer}}]{mazoyer_cortical_2001}%
  \BibitemOpen
  \bibfield  {author} {\bibinfo {author} {\bibfnamefont {B.}~\bibnamefont
  {Mazoyer}}, \bibinfo {author} {\bibfnamefont {L.}~\bibnamefont {Zago}},
  \bibinfo {author} {\bibfnamefont {E.}~\bibnamefont {Mellet}}, \bibinfo
  {author} {\bibfnamefont {S.}~\bibnamefont {Bricogne}}, \bibinfo {author}
  {\bibfnamefont {O.}~\bibnamefont {Etard}}, \bibinfo {author} {\bibfnamefont
  {O.}~\bibnamefont {Houd{\'e}}}, \bibinfo {author} {\bibfnamefont
  {F.}~\bibnamefont {Crivello}}, \bibinfo {author} {\bibfnamefont
  {M.}~\bibnamefont {Joliot}}, \bibinfo {author} {\bibfnamefont
  {L.}~\bibnamefont {Petit}}, \ and\ \bibinfo {author} {\bibfnamefont
  {N.}~\bibnamefont {Tzourio-Mazoyer}},\ }\bibfield  {title} {\enquote
  {\bibinfo {title} {Cortical networks for working memory and executive
  functions sustain the conscious resting state in man},}\ }\href@noop {}
  {\bibfield  {journal} {\bibinfo  {journal} {Brain Res. Bull.}\ }\textbf
  {\bibinfo {volume} {54}},\ \bibinfo {pages} {287--298} (\bibinfo {year}
  {2001})}\BibitemShut {NoStop}%
\bibitem [{\citenamefont {Shulman}\ \emph {et~al.}(2004)\citenamefont
  {Shulman}, \citenamefont {Rothman}, \citenamefont {Behar},\ and\
  \citenamefont {Hyder}}]{shulman_energetic_2004}%
  \BibitemOpen
  \bibfield  {author} {\bibinfo {author} {\bibfnamefont {R.~G.}\ \bibnamefont
  {Shulman}}, \bibinfo {author} {\bibfnamefont {D.~L.}\ \bibnamefont
  {Rothman}}, \bibinfo {author} {\bibfnamefont {K.~L.}\ \bibnamefont {Behar}},
  \ and\ \bibinfo {author} {\bibfnamefont {F.}~\bibnamefont {Hyder}},\
  }\bibfield  {title} {\enquote {\bibinfo {title} {Energetic basis of brain
  activity: implications for neuroimaging},}\ }\href@noop {} {\bibfield
  {journal} {\bibinfo  {journal} {Trends Neurosci.}\ }\textbf {\bibinfo
  {volume} {27}},\ \bibinfo {pages} {489--495} (\bibinfo {year}
  {2004})}\BibitemShut {NoStop}%
\bibitem [{\citenamefont {Fox}\ \emph {et~al.}(2005)\citenamefont {Fox},
  \citenamefont {Snyder}, \citenamefont {Vincent}, \citenamefont {Corbetta},
  \citenamefont {Essen},\ and\ \citenamefont {Raichle}}]{fox_human_2005}%
  \BibitemOpen
  \bibfield  {author} {\bibinfo {author} {\bibfnamefont {M.~D.}\ \bibnamefont
  {Fox}}, \bibinfo {author} {\bibfnamefont {A.~Z.}\ \bibnamefont {Snyder}},
  \bibinfo {author} {\bibfnamefont {J.~L.}\ \bibnamefont {Vincent}}, \bibinfo
  {author} {\bibfnamefont {M.}~\bibnamefont {Corbetta}}, \bibinfo {author}
  {\bibfnamefont {D.~C.~V.}\ \bibnamefont {Essen}}, \ and\ \bibinfo {author}
  {\bibfnamefont {M.~E.}\ \bibnamefont {Raichle}},\ }\bibfield  {title}
  {\enquote {\bibinfo {title} {The human brain is intrinsically organized into
  dynamic, anticorrelated functional networks},}\ }\href@noop {} {\bibfield
  {journal} {\bibinfo  {journal} {Proc. Nat. Acad. Sci. USA}\ }\textbf
  {\bibinfo {volume} {102}},\ \bibinfo {pages} {9673--9678} (\bibinfo {year}
  {2005})}\BibitemShut {NoStop}%
\bibitem [{\citenamefont {Beckmann}\ \emph {et~al.}(2005)\citenamefont
  {Beckmann}, \citenamefont {DeLuca}, \citenamefont {Devlin},\ and\
  \citenamefont {Smith}}]{beckmann_investigations_2005}%
  \BibitemOpen
  \bibfield  {author} {\bibinfo {author} {\bibfnamefont {C.~F.}\ \bibnamefont
  {Beckmann}}, \bibinfo {author} {\bibfnamefont {M.}~\bibnamefont {DeLuca}},
  \bibinfo {author} {\bibfnamefont {J.~T.}\ \bibnamefont {Devlin}}, \ and\
  \bibinfo {author} {\bibfnamefont {S.~M.}\ \bibnamefont {Smith}},\ }\bibfield
  {title} {\enquote {\bibinfo {title} {Investigations into resting-state
  connectivity using independent component analysis},}\ }\href@noop {}
  {\bibfield  {journal} {\bibinfo  {journal} {Phil. Trans. R. Soc. B}\ }\textbf
  {\bibinfo {volume} {360}},\ \bibinfo {pages} {1001--1013} (\bibinfo {year}
  {2005})}\BibitemShut {NoStop}%
\bibitem [{\citenamefont {Fransson}(2006)}]{fransson_how_2006}%
  \BibitemOpen
  \bibfield  {author} {\bibinfo {author} {\bibfnamefont {P.}~\bibnamefont
  {Fransson}},\ }\bibfield  {title} {\enquote {\bibinfo {title} {How default is
  the default mode of brain function?: {Further} evidence from intrinsic {BOLD}
  signal fluctuations},}\ }\href@noop {} {\bibfield  {journal} {\bibinfo
  {journal} {Neuropsychologia}\ }\textbf {\bibinfo {volume} {44}},\ \bibinfo
  {pages} {2836--2845} (\bibinfo {year} {2006})}\BibitemShut {NoStop}%
\bibitem [{\citenamefont {Smith}\ \emph {et~al.}(2009)\citenamefont {Smith},
  \citenamefont {Fox}, \citenamefont {Miller}, \citenamefont {Glahn},
  \citenamefont {Fox}, \citenamefont {Mackay}, \citenamefont {Filippini},
  \citenamefont {Watkins}, \citenamefont {Toro}, \citenamefont {Laird},\ and\
  \citenamefont {Beckmann}}]{smith_correspondence_2009}%
  \BibitemOpen
  \bibfield  {author} {\bibinfo {author} {\bibfnamefont {S.~M.}\ \bibnamefont
  {Smith}}, \bibinfo {author} {\bibfnamefont {P.~T.}\ \bibnamefont {Fox}},
  \bibinfo {author} {\bibfnamefont {K.~L.}\ \bibnamefont {Miller}}, \bibinfo
  {author} {\bibfnamefont {D.~C.}\ \bibnamefont {Glahn}}, \bibinfo {author}
  {\bibfnamefont {P.~M.}\ \bibnamefont {Fox}}, \bibinfo {author} {\bibfnamefont
  {C.~E.}\ \bibnamefont {Mackay}}, \bibinfo {author} {\bibfnamefont
  {N.}~\bibnamefont {Filippini}}, \bibinfo {author} {\bibfnamefont {K.~E.}\
  \bibnamefont {Watkins}}, \bibinfo {author} {\bibfnamefont {R.}~\bibnamefont
  {Toro}}, \bibinfo {author} {\bibfnamefont {A.~R.}\ \bibnamefont {Laird}}, \
  and\ \bibinfo {author} {\bibfnamefont {C.~F.}\ \bibnamefont {Beckmann}},\
  }\bibfield  {title} {\enquote {\bibinfo {title} {Correspondence of the
  brain's functional architecture during activation and rest},}\ }\href@noop {}
  {\bibfield  {journal} {\bibinfo  {journal} {Proc. Nat. Acad. Sci. USA}\
  }\textbf {\bibinfo {volume} {106}},\ \bibinfo {pages} {13040--13045}
  (\bibinfo {year} {2009})}\BibitemShut {NoStop}%
\bibitem [{\citenamefont {Biswal}\ \emph {et~al.}(2010)\citenamefont {Biswal},
  \citenamefont {Mennes}, \citenamefont {Zuo}, \citenamefont {Gohel},
  \citenamefont {Kelly}, \citenamefont {Smith}, \citenamefont {Beckmann},
  \citenamefont {Adelstein}, \citenamefont {Buckner}, \citenamefont {Colcombe},
  \citenamefont {Dogonowski}, \citenamefont {Ernst}, \citenamefont {Fair},
  \citenamefont {Hampson}, \citenamefont {Hoptman}, \citenamefont {Hyde},
  \citenamefont {Kiviniemi}, \citenamefont {K{\"o}tter}, \citenamefont {Li},
  \citenamefont {Lin}, \citenamefont {Lowe}, \citenamefont {Mackay},
  \citenamefont {Madden}, \citenamefont {Madsen}, \citenamefont {Margulies},
  \citenamefont {Mayberg}, \citenamefont {McMahon}, \citenamefont {Monk},
  \citenamefont {Mostofsky}, \citenamefont {Nagel}, \citenamefont {Pekar},
  \citenamefont {Peltier}, \citenamefont {Petersen}, \citenamefont {Riedl},
  \citenamefont {Rombouts}, \citenamefont {Rypma}, \citenamefont {Schlaggar},
  \citenamefont {Schmidt}, \citenamefont {Seidler}, \citenamefont {Siegle},
  \citenamefont {Sorg}, \citenamefont {Teng}, \citenamefont {Veijola},
  \citenamefont {Villringer}, \citenamefont {Walter}, \citenamefont {Wang},
  \citenamefont {Weng}, \citenamefont {Whitfield-Gabrieli}, \citenamefont
  {Williamson}, \citenamefont {Windischberger}, \citenamefont {Zang},
  \citenamefont {Zhang}, \citenamefont {Castellanos},\ and\ \citenamefont
  {Milham}}]{biswal_toward_2010}%
  \BibitemOpen
  \bibfield  {author} {\bibinfo {author} {\bibfnamefont {B.~B.}\ \bibnamefont
  {Biswal}}, \bibinfo {author} {\bibfnamefont {M.}~\bibnamefont {Mennes}},
  \bibinfo {author} {\bibfnamefont {X.-N.}\ \bibnamefont {Zuo}}, \bibinfo
  {author} {\bibfnamefont {S.}~\bibnamefont {Gohel}}, \bibinfo {author}
  {\bibfnamefont {C.}~\bibnamefont {Kelly}}, \bibinfo {author} {\bibfnamefont
  {S.~M.}\ \bibnamefont {Smith}}, \bibinfo {author} {\bibfnamefont {C.~F.}\
  \bibnamefont {Beckmann}}, \bibinfo {author} {\bibfnamefont {J.~S.}\
  \bibnamefont {Adelstein}}, \bibinfo {author} {\bibfnamefont {R.~L.}\
  \bibnamefont {Buckner}}, \bibinfo {author} {\bibfnamefont {S.}~\bibnamefont
  {Colcombe}}, \bibinfo {author} {\bibfnamefont {A.-M.}\ \bibnamefont
  {Dogonowski}}, \bibinfo {author} {\bibfnamefont {M.}~\bibnamefont {Ernst}},
  \bibinfo {author} {\bibfnamefont {D.}~\bibnamefont {Fair}}, \bibinfo {author}
  {\bibfnamefont {M.}~\bibnamefont {Hampson}}, \bibinfo {author} {\bibfnamefont
  {M.~J.}\ \bibnamefont {Hoptman}}, \bibinfo {author} {\bibfnamefont {J.~S.}\
  \bibnamefont {Hyde}}, \bibinfo {author} {\bibfnamefont {V.~J.}\ \bibnamefont
  {Kiviniemi}}, \bibinfo {author} {\bibfnamefont {R.}~\bibnamefont
  {K{\"o}tter}}, \bibinfo {author} {\bibfnamefont {S.-J.}\ \bibnamefont {Li}},
  \bibinfo {author} {\bibfnamefont {C.-P.}\ \bibnamefont {Lin}}, \bibinfo
  {author} {\bibfnamefont {M.~J.}\ \bibnamefont {Lowe}}, \bibinfo {author}
  {\bibfnamefont {C.}~\bibnamefont {Mackay}}, \bibinfo {author} {\bibfnamefont
  {D.~J.}\ \bibnamefont {Madden}}, \bibinfo {author} {\bibfnamefont {K.~H.}\
  \bibnamefont {Madsen}}, \bibinfo {author} {\bibfnamefont {D.~S.}\
  \bibnamefont {Margulies}}, \bibinfo {author} {\bibfnamefont {H.~S.}\
  \bibnamefont {Mayberg}}, \bibinfo {author} {\bibfnamefont {K.}~\bibnamefont
  {McMahon}}, \bibinfo {author} {\bibfnamefont {C.~S.}\ \bibnamefont {Monk}},
  \bibinfo {author} {\bibfnamefont {S.~H.}\ \bibnamefont {Mostofsky}}, \bibinfo
  {author} {\bibfnamefont {B.~J.}\ \bibnamefont {Nagel}}, \bibinfo {author}
  {\bibfnamefont {J.~J.}\ \bibnamefont {Pekar}}, \bibinfo {author}
  {\bibfnamefont {S.~J.}\ \bibnamefont {Peltier}}, \bibinfo {author}
  {\bibfnamefont {S.~E.}\ \bibnamefont {Petersen}}, \bibinfo {author}
  {\bibfnamefont {V.}~\bibnamefont {Riedl}}, \bibinfo {author} {\bibfnamefont
  {S.~A. R.~B.}\ \bibnamefont {Rombouts}}, \bibinfo {author} {\bibfnamefont
  {B.}~\bibnamefont {Rypma}}, \bibinfo {author} {\bibfnamefont {B.~L.}\
  \bibnamefont {Schlaggar}}, \bibinfo {author} {\bibfnamefont {S.}~\bibnamefont
  {Schmidt}}, \bibinfo {author} {\bibfnamefont {R.~D.}\ \bibnamefont
  {Seidler}}, \bibinfo {author} {\bibfnamefont {G.~J.}\ \bibnamefont {Siegle}},
  \bibinfo {author} {\bibfnamefont {C.}~\bibnamefont {Sorg}}, \bibinfo {author}
  {\bibfnamefont {G.-J.}\ \bibnamefont {Teng}}, \bibinfo {author}
  {\bibfnamefont {J.}~\bibnamefont {Veijola}}, \bibinfo {author} {\bibfnamefont
  {A.}~\bibnamefont {Villringer}}, \bibinfo {author} {\bibfnamefont
  {M.}~\bibnamefont {Walter}}, \bibinfo {author} {\bibfnamefont
  {L.}~\bibnamefont {Wang}}, \bibinfo {author} {\bibfnamefont {X.-C.}\
  \bibnamefont {Weng}}, \bibinfo {author} {\bibfnamefont {S.}~\bibnamefont
  {Whitfield-Gabrieli}}, \bibinfo {author} {\bibfnamefont {P.}~\bibnamefont
  {Williamson}}, \bibinfo {author} {\bibfnamefont {C.}~\bibnamefont
  {Windischberger}}, \bibinfo {author} {\bibfnamefont {Y.-F.}\ \bibnamefont
  {Zang}}, \bibinfo {author} {\bibfnamefont {H.-Y.}\ \bibnamefont {Zhang}},
  \bibinfo {author} {\bibfnamefont {F.~X.}\ \bibnamefont {Castellanos}}, \ and\
  \bibinfo {author} {\bibfnamefont {M.~P.}\ \bibnamefont {Milham}},\ }\bibfield
   {title} {\enquote {\bibinfo {title} {Toward discovery science of human brain
  function},}\ }\href@noop {} {\bibfield  {journal} {\bibinfo  {journal} {Proc.
  Nat. Acad. Sci. USA}\ }\textbf {\bibinfo {volume} {107}},\ \bibinfo {pages}
  {4734--4739} (\bibinfo {year} {2010})}\BibitemShut {NoStop}%
\bibitem [{\citenamefont {Laufs}\ \emph {et~al.}(2003)\citenamefont {Laufs},
  \citenamefont {Krakow}, \citenamefont {Sterzer}, \citenamefont {Eger},
  \citenamefont {Beyerle}, \citenamefont {Salek-Haddadi},\ and\ \citenamefont
  {Kleinschmidt}}]{laufs_electroencephalographic_2003}%
  \BibitemOpen
  \bibfield  {author} {\bibinfo {author} {\bibfnamefont {H.}~\bibnamefont
  {Laufs}}, \bibinfo {author} {\bibfnamefont {K.}~\bibnamefont {Krakow}},
  \bibinfo {author} {\bibfnamefont {P.}~\bibnamefont {Sterzer}}, \bibinfo
  {author} {\bibfnamefont {E.}~\bibnamefont {Eger}}, \bibinfo {author}
  {\bibfnamefont {A.}~\bibnamefont {Beyerle}}, \bibinfo {author} {\bibfnamefont
  {A.}~\bibnamefont {Salek-Haddadi}}, \ and\ \bibinfo {author} {\bibfnamefont
  {A.}~\bibnamefont {Kleinschmidt}},\ }\bibfield  {title} {\enquote {\bibinfo
  {title} {Electroencephalographic signatures of attentional and cognitive
  default modes in spontaneous brain activity fluctuations at rest},}\
  }\href@noop {} {\bibfield  {journal} {\bibinfo  {journal} {Proc. Nat. Acad.
  Sci. USA}\ }\textbf {\bibinfo {volume} {100}},\ \bibinfo {pages}
  {11053--11058} (\bibinfo {year} {2003})}\BibitemShut {NoStop}%
\bibitem [{\citenamefont {Brookes}\ \emph {et~al.}(2011)\citenamefont
  {Brookes}, \citenamefont {Woolrich}, \citenamefont {Luckhoo}, \citenamefont
  {Price}, \citenamefont {Hale}, \citenamefont {Stephenson}, \citenamefont
  {Barnes}, \citenamefont {Smith},\ and\ \citenamefont
  {Morris}}]{brookes2011investigating}%
  \BibitemOpen
  \bibfield  {author} {\bibinfo {author} {\bibfnamefont {M.~J.}\ \bibnamefont
  {Brookes}}, \bibinfo {author} {\bibfnamefont {M.}~\bibnamefont {Woolrich}},
  \bibinfo {author} {\bibfnamefont {H.}~\bibnamefont {Luckhoo}}, \bibinfo
  {author} {\bibfnamefont {D.}~\bibnamefont {Price}}, \bibinfo {author}
  {\bibfnamefont {J.~R.}\ \bibnamefont {Hale}}, \bibinfo {author}
  {\bibfnamefont {M.~C.}\ \bibnamefont {Stephenson}}, \bibinfo {author}
  {\bibfnamefont {G.~R.}\ \bibnamefont {Barnes}}, \bibinfo {author}
  {\bibfnamefont {S.~M.}\ \bibnamefont {Smith}}, \ and\ \bibinfo {author}
  {\bibfnamefont {P.~G.}\ \bibnamefont {Morris}},\ }\bibfield  {title}
  {\enquote {\bibinfo {title} {Investigating the electrophysiological basis of
  resting state networks using magnetoencephalography},}\ }\href@noop {}
  {\bibfield  {journal} {\bibinfo  {journal} {Proc. Nat. Acad. Sci. USA}\
  }\textbf {\bibinfo {volume} {108}},\ \bibinfo {pages} {16783--16788}
  (\bibinfo {year} {2011})}\BibitemShut {NoStop}%
\bibitem [{\citenamefont {Ponce-Alvarez}\ \emph {et~al.}(2015)\citenamefont
  {Ponce-Alvarez}, \citenamefont {Deco}, \citenamefont {Hagmann}, \citenamefont
  {Romani}, \citenamefont {Mantini},\ and\ \citenamefont
  {Corbetta}}]{ponce-alvarez_resting-state_2015}%
  \BibitemOpen
  \bibfield  {author} {\bibinfo {author} {\bibfnamefont {A.}~\bibnamefont
  {Ponce-Alvarez}}, \bibinfo {author} {\bibfnamefont {G.}~\bibnamefont {Deco}},
  \bibinfo {author} {\bibfnamefont {P.}~\bibnamefont {Hagmann}}, \bibinfo
  {author} {\bibfnamefont {G.~L.}\ \bibnamefont {Romani}}, \bibinfo {author}
  {\bibfnamefont {D.}~\bibnamefont {Mantini}}, \ and\ \bibinfo {author}
  {\bibfnamefont {M.}~\bibnamefont {Corbetta}},\ }\bibfield  {title} {\enquote
  {\bibinfo {title} {Resting-{State} {Temporal} {Synchronization} {Networks}
  {Emerge} from {Connectivity} {Topology} and {Heterogeneity}},}\ }\href@noop
  {} {\bibfield  {journal} {\bibinfo  {journal} {PLOS Comput Biol}\ }\textbf
  {\bibinfo {volume} {11}},\ \bibinfo {pages} {e1004100} (\bibinfo {year}
  {2015})}\BibitemShut {NoStop}%
\bibitem [{\citenamefont {Keilholz}\ \emph {et~al.}(2016)\citenamefont
  {Keilholz}, \citenamefont {Billings}, \citenamefont {Wang}, \citenamefont
  {Abbas}, \citenamefont {Hafeneger}, \citenamefont {Pan}, \citenamefont
  {Shakil},\ and\ \citenamefont {Nezafati}}]{keilholz2016multiscale}%
  \BibitemOpen
  \bibfield  {author} {\bibinfo {author} {\bibfnamefont {S.~D.}\ \bibnamefont
  {Keilholz}}, \bibinfo {author} {\bibfnamefont {J.~C.~W.}\ \bibnamefont
  {Billings}}, \bibinfo {author} {\bibfnamefont {K.}~\bibnamefont {Wang}},
  \bibinfo {author} {\bibfnamefont {A.}~\bibnamefont {Abbas}}, \bibinfo
  {author} {\bibfnamefont {C.}~\bibnamefont {Hafeneger}}, \bibinfo {author}
  {\bibfnamefont {W.~J.}\ \bibnamefont {Pan}}, \bibinfo {author} {\bibfnamefont
  {S.}~\bibnamefont {Shakil}}, \ and\ \bibinfo {author} {\bibfnamefont
  {M.}~\bibnamefont {Nezafati}},\ }\bibfield  {title} {\enquote {\bibinfo
  {title} {Multiscale network activity in resting state {fMRI}},}\ }in\
  \href@noop {} {\emph {\bibinfo {booktitle} {2016 38th {Annual}
  {International} {Conference} of the {IEEE} {Engineering} in {Medicine} and
  {Biology} {Society} ({EMBC})}}}\ (\bibinfo {year} {2016})\ pp.\ \bibinfo
  {pages} {61--64}\BibitemShut {NoStop}%
\bibitem [{\citenamefont {Deco}\ \emph {et~al.}(2013)\citenamefont {Deco},
  \citenamefont {Hagmann}, \citenamefont {Hudetz},\ and\ \citenamefont
  {Tononi}}]{Deco_CortexAsleep_2013}%
  \BibitemOpen
  \bibfield  {author} {\bibinfo {author} {\bibfnamefont {G.}~\bibnamefont
  {Deco}}, \bibinfo {author} {\bibfnamefont {P.}~\bibnamefont {Hagmann}},
  \bibinfo {author} {\bibfnamefont {A.}~\bibnamefont {Hudetz}}, \ and\ \bibinfo
  {author} {\bibfnamefont {G.}~\bibnamefont {Tononi}},\ }\bibfield  {title}
  {\enquote {\bibinfo {title} {Modeling resting-state functional networks when
  the cortex falls asleep: Local and global changes},}\ }\href@noop {}
  {\bibfield  {journal} {\bibinfo  {journal} {Cereb. Cortex}\ }\textbf
  {\bibinfo {volume} {24}},\ \bibinfo {pages} {3180--3194} (\bibinfo {year}
  {2013})}\BibitemShut {NoStop}%
\bibitem [{\citenamefont {Hudetz}, \citenamefont {Humphries},\ and\
  \citenamefont {Binder}(2014)}]{Hudetz_SpinGlass_2014}%
  \BibitemOpen
  \bibfield  {author} {\bibinfo {author} {\bibfnamefont {A.}~\bibnamefont
  {Hudetz}}, \bibinfo {author} {\bibfnamefont {C.}~\bibnamefont {Humphries}}, \
  and\ \bibinfo {author} {\bibfnamefont {J.}~\bibnamefont {Binder}},\
  }\bibfield  {title} {\enquote {\bibinfo {title} {Spin-glass model predicts
  metastable brain states that diminish in anesthesia},}\ }\href@noop {}
  {\bibfield  {journal} {\bibinfo  {journal} {Front. Syst. Neurosci.}\ }\textbf
  {\bibinfo {volume} {8}},\ \bibinfo {pages} {234} (\bibinfo {year}
  {2014})}\BibitemShut {NoStop}%
\bibitem [{\citenamefont {Hudetz}, \citenamefont {Liu},\ and\ \citenamefont
  {Pillay}(2014)}]{Hudetz_Unconsciousness_2014}%
  \BibitemOpen
  \bibfield  {author} {\bibinfo {author} {\bibfnamefont {A.}~\bibnamefont
  {Hudetz}}, \bibinfo {author} {\bibfnamefont {X.}~\bibnamefont {Liu}}, \ and\
  \bibinfo {author} {\bibfnamefont {S.}~\bibnamefont {Pillay}},\ }\bibfield
  {title} {\enquote {\bibinfo {title} {Dynamic repertoire of intrinsic brain
  states is reduced in propofol-induced unconsciousness},}\ }\href@noop {}
  {\bibfield  {journal} {\bibinfo  {journal} {Brain Connectivity}\ }\textbf
  {\bibinfo {volume} {0}},\ \bibinfo {pages} {1--13} (\bibinfo {year}
  {2014})}\BibitemShut {NoStop}%
\bibitem [{\citenamefont {Bettinardi}\ \emph {et~al.}(2015)\citenamefont
  {Bettinardi}, \citenamefont {Tort-Colet}, \citenamefont {Ruiz-Mejias},
  \citenamefont {Sanchez-Vives},\ and\ \citenamefont
  {Deco}}]{bettinardi_gradual_2015}%
  \BibitemOpen
  \bibfield  {author} {\bibinfo {author} {\bibfnamefont {R.~G.}\ \bibnamefont
  {Bettinardi}}, \bibinfo {author} {\bibfnamefont {N.}~\bibnamefont
  {Tort-Colet}}, \bibinfo {author} {\bibfnamefont {M.}~\bibnamefont
  {Ruiz-Mejias}}, \bibinfo {author} {\bibfnamefont {M.~V.}\ \bibnamefont
  {Sanchez-Vives}}, \ and\ \bibinfo {author} {\bibfnamefont {G.}~\bibnamefont
  {Deco}},\ }\bibfield  {title} {\enquote {\bibinfo {title} {Gradual emergence
  of spontaneous correlated brain activity during fading of general anesthesia
  in rats: {Evidences} from {fMRI} and local field potentials},}\ }\href@noop
  {} {\bibfield  {journal} {\bibinfo  {journal} {NeuroImage}\ }\textbf
  {\bibinfo {volume} {114}},\ \bibinfo {pages} {185--198} (\bibinfo {year}
  {2015})}\BibitemShut {NoStop}%
\bibitem [{\citenamefont {Zhou}\ \emph {et~al.}(2006)\citenamefont {Zhou},
  \citenamefont {Zemanov\'a}, \citenamefont {Zamora-L\'opez}, \citenamefont
  {Hilgetag},\ and\ \citenamefont {Kurths}}]{Zhou_HierarchicalCat_2006}%
  \BibitemOpen
  \bibfield  {author} {\bibinfo {author} {\bibfnamefont {C.~S.}\ \bibnamefont
  {Zhou}}, \bibinfo {author} {\bibfnamefont {L.}~\bibnamefont {Zemanov\'a}},
  \bibinfo {author} {\bibfnamefont {G.}~\bibnamefont {Zamora-L\'opez}},
  \bibinfo {author} {\bibfnamefont {C.-C.}\ \bibnamefont {Hilgetag}}, \ and\
  \bibinfo {author} {\bibfnamefont {J.}~\bibnamefont {Kurths}},\ }\bibfield
  {title} {\enquote {\bibinfo {title} {Hierarchical organization unveiled by
  functional connectivity in complex brain networks},}\ }\href@noop {}
  {\bibfield  {journal} {\bibinfo  {journal} {Phys. Rev. Lett.}\ }\textbf
  {\bibinfo {volume} {97}},\ \bibinfo {pages} {238103} (\bibinfo {year}
  {2006})}\BibitemShut {NoStop}%
\bibitem [{\citenamefont {Honey}\ \emph {et~al.}(2007)\citenamefont {Honey},
  \citenamefont {Kotter}, \citenamefont {Breakspear},\ and\ \citenamefont
  {Sporns}}]{honey_network_2007}%
  \BibitemOpen
  \bibfield  {author} {\bibinfo {author} {\bibfnamefont {C.~J.}\ \bibnamefont
  {Honey}}, \bibinfo {author} {\bibfnamefont {R.}~\bibnamefont {Kotter}},
  \bibinfo {author} {\bibfnamefont {M.}~\bibnamefont {Breakspear}}, \ and\
  \bibinfo {author} {\bibfnamefont {O.}~\bibnamefont {Sporns}},\ }\bibfield
  {title} {\enquote {\bibinfo {title} {Network structure of cerebral cortex
  shapes functional connectivity on multiple time scales},}\ }\href@noop {}
  {\bibfield  {journal} {\bibinfo  {journal} {Proc. Nat. Acad. Sci. USA}\
  }\textbf {\bibinfo {volume} {104}},\ \bibinfo {pages} {10240--10245}
  (\bibinfo {year} {2007})}\BibitemShut {NoStop}%
\bibitem [{\citenamefont {G\'omez-Garde{\~n}es}\ \emph
  {et~al.}(2010)\citenamefont {G\'omez-Garde{\~n}es}, \citenamefont
  {Zamora-L{\'o}pez}, \citenamefont {Moreno},\ and\ \citenamefont
  {Arenas}}]{Gomez_FromModular_2010}%
  \BibitemOpen
  \bibfield  {author} {\bibinfo {author} {\bibfnamefont {J.}~\bibnamefont
  {G\'omez-Garde{\~n}es}}, \bibinfo {author} {\bibfnamefont {G.}~\bibnamefont
  {Zamora-L{\'o}pez}}, \bibinfo {author} {\bibfnamefont {Y.}~\bibnamefont
  {Moreno}}, \ and\ \bibinfo {author} {\bibfnamefont {A.}~\bibnamefont
  {Arenas}},\ }\bibfield  {title} {\enquote {\bibinfo {title} {From modular to
  centralized organization of synchronization in functional areas of the cat
  cerebral cortex},}\ }\href@noop {} {\bibfield  {journal} {\bibinfo  {journal}
  {PLoS ONE}\ }\textbf {\bibinfo {volume} {5}},\ \bibinfo {pages} {e12313}
  (\bibinfo {year} {2010})}\BibitemShut {NoStop}%
\bibitem [{\citenamefont {Deco}\ \emph {et~al.}(2009)\citenamefont {Deco},
  \citenamefont {Jirsa}, \citenamefont {McIntosh}, \citenamefont {Sporns},\
  and\ \citenamefont {Kotter}}]{deco_key_2009}%
  \BibitemOpen
  \bibfield  {author} {\bibinfo {author} {\bibfnamefont {G.}~\bibnamefont
  {Deco}}, \bibinfo {author} {\bibfnamefont {V.}~\bibnamefont {Jirsa}},
  \bibinfo {author} {\bibfnamefont {A.~R.}\ \bibnamefont {McIntosh}}, \bibinfo
  {author} {\bibfnamefont {O.}~\bibnamefont {Sporns}}, \ and\ \bibinfo {author}
  {\bibfnamefont {R.}~\bibnamefont {Kotter}},\ }\bibfield  {title} {\enquote
  {\bibinfo {title} {Key role of coupling, delay, and noise in resting brain
  fluctuations},}\ }\href@noop {} {\bibfield  {journal} {\bibinfo  {journal}
  {Proc. Nat. Acad. Sci. USA}\ }\textbf {\bibinfo {volume} {106}},\ \bibinfo
  {pages} {10302--10307} (\bibinfo {year} {2009})}\BibitemShut {NoStop}%
\bibitem [{\citenamefont {Deco}, \citenamefont {Jirsa},\ and\ \citenamefont
  {McIntosh}(2011)}]{deco_emerging_2011}%
  \BibitemOpen
  \bibfield  {author} {\bibinfo {author} {\bibfnamefont {G.}~\bibnamefont
  {Deco}}, \bibinfo {author} {\bibfnamefont {V.~K.}\ \bibnamefont {Jirsa}}, \
  and\ \bibinfo {author} {\bibfnamefont {A.~R.}\ \bibnamefont {McIntosh}},\
  }\bibfield  {title} {\enquote {\bibinfo {title} {Emerging concepts for the
  dynamical organization of resting-state activity in the brain},}\ }\href@noop
  {} {\bibfield  {journal} {\bibinfo  {journal} {Nat. Rev. Neurosci.}\ }\textbf
  {\bibinfo {volume} {12}},\ \bibinfo {pages} {43--56} (\bibinfo {year}
  {2011})}\BibitemShut {NoStop}%
\bibitem [{\citenamefont {Deco}, \citenamefont {Jirsa},\ and\ \citenamefont
  {McIntosh}(2013)}]{deco_resting_2013}%
  \BibitemOpen
  \bibfield  {author} {\bibinfo {author} {\bibfnamefont {G.}~\bibnamefont
  {Deco}}, \bibinfo {author} {\bibfnamefont {V.~K.}\ \bibnamefont {Jirsa}}, \
  and\ \bibinfo {author} {\bibfnamefont {A.~R.}\ \bibnamefont {McIntosh}},\
  }\bibfield  {title} {\enquote {\bibinfo {title} {Resting brains never rest:
  computational insights into potential cognitive architectures},}\ }\href@noop
  {} {\bibfield  {journal} {\bibinfo  {journal} {Trends Neurosci.}\ }\textbf
  {\bibinfo {volume} {36}},\ \bibinfo {pages} {268--274} (\bibinfo {year}
  {2013})}\BibitemShut {NoStop}%
\bibitem [{\citenamefont {Deco}\ and\ \citenamefont
  {Kringelbach}(2014)}]{deco_great_2014}%
  \BibitemOpen
  \bibfield  {author} {\bibinfo {author} {\bibfnamefont {G.}~\bibnamefont
  {Deco}}\ and\ \bibinfo {author} {\bibfnamefont {M.~L.}\ \bibnamefont
  {Kringelbach}},\ }\bibfield  {title} {\enquote {\bibinfo {title} {Great
  expectations: {Using} whole-brain computational connectomics for
  understanding neuropsychiatric disorders},}\ }\href@noop {} {\bibfield
  {journal} {\bibinfo  {journal} {Neuron}\ }\textbf {\bibinfo {volume} {84}},\
  \bibinfo {pages} {892--905} (\bibinfo {year} {2014})}\BibitemShut {NoStop}%
\bibitem [{\citenamefont {Mess\'e}\ \emph {et~al.}(2014)\citenamefont
  {Mess\'e}, \citenamefont {Rudrauf}, \citenamefont {Benali},\ and\
  \citenamefont {Marrelec}}]{messe_relating_2014}%
  \BibitemOpen
  \bibfield  {author} {\bibinfo {author} {\bibfnamefont {A.}~\bibnamefont
  {Mess\'e}}, \bibinfo {author} {\bibfnamefont {D.}~\bibnamefont {Rudrauf}},
  \bibinfo {author} {\bibfnamefont {H.}~\bibnamefont {Benali}}, \ and\ \bibinfo
  {author} {\bibfnamefont {G.}~\bibnamefont {Marrelec}},\ }\bibfield  {title}
  {\enquote {\bibinfo {title} {Relating structure and function in the human
  brain: relative contributions of anatomy, stationary dynamics, and
  non-stationarities},}\ }\href@noop {} {\bibfield  {journal} {\bibinfo
  {journal} {PLoS Comput. Biol.}\ }\textbf {\bibinfo {volume} {10}},\ \bibinfo
  {pages} {e1003530} (\bibinfo {year} {2014})}\BibitemShut {NoStop}%
\bibitem [{\citenamefont {Schmidt}\ \emph {et~al.}(2010)\citenamefont
  {Schmidt}, \citenamefont {Zamora-L\'opez}, \citenamefont {Zhou},\ and\
  \citenamefont {Kurths}}]{schmidt_simulation_2010}%
  \BibitemOpen
  \bibfield  {author} {\bibinfo {author} {\bibfnamefont {G.}~\bibnamefont
  {Schmidt}}, \bibinfo {author} {\bibfnamefont {G.}~\bibnamefont
  {Zamora-L\'opez}}, \bibinfo {author} {\bibfnamefont {C.}~\bibnamefont
  {Zhou}}, \ and\ \bibinfo {author} {\bibfnamefont {J.}~\bibnamefont
  {Kurths}},\ }\bibfield  {title} {\enquote {\bibinfo {title} {Simulation of
  large scale cortical networks by individual neuron dynamics},}\ }\href@noop
  {} {\bibfield  {journal} {\bibinfo  {journal} {Int. J. Bif. \& Chaos}\
  }\textbf {\bibinfo {volume} {20}},\ \bibinfo {pages} {859--867} (\bibinfo
  {year} {2010})}\BibitemShut {NoStop}%
\bibitem [{\citenamefont {Latora}\ and\ \citenamefont
  {Marchiori}(2001)}]{latora_efficient_2001}%
  \BibitemOpen
  \bibfield  {author} {\bibinfo {author} {\bibfnamefont {V.}~\bibnamefont
  {Latora}}\ and\ \bibinfo {author} {\bibfnamefont {M.}~\bibnamefont
  {Marchiori}},\ }\bibfield  {title} {\enquote {\bibinfo {title} {Efficient
  behavior of small-world networks},}\ }\href@noop {} {\bibfield  {journal}
  {\bibinfo  {journal} {Phys. Rev. Lett.}\ }\textbf {\bibinfo {volume} {87}},\
  \bibinfo {pages} {198701} (\bibinfo {year} {2001})}\BibitemShut {NoStop}%
\bibitem [{\citenamefont {Huberman}\ and\ \citenamefont
  {Adamic}(2004)}]{huberman_information_2004}%
  \BibitemOpen
  \bibfield  {author} {\bibinfo {author} {\bibfnamefont {B.~A.}\ \bibnamefont
  {Huberman}}\ and\ \bibinfo {author} {\bibfnamefont {L.~A.}\ \bibnamefont
  {Adamic}},\ }\bibfield  {title} {\enquote {\bibinfo {title} {Information
  dynamics in the networked world},}\ }in\ \href@noop {} {\emph {\bibinfo
  {booktitle} {Complex {Networks}}}},\ \bibinfo {series and number} {\bibinfo
  {series} {Lecture {Notes} in {Physics}}\ No.\ \bibinfo {number} {650}},\
  \bibinfo {editor} {edited by\ \bibinfo {editor} {\bibfnamefont
  {E.}~\bibnamefont {Ben-Naim}}, \bibinfo {editor} {\bibfnamefont
  {H.}~\bibnamefont {Frauenfelder}}, \ and\ \bibinfo {editor} {\bibfnamefont
  {Z.}~\bibnamefont {Toroczkai}}}\ (\bibinfo  {publisher} {Springer Berlin
  Heidelberg},\ \bibinfo {year} {2004})\ pp.\ \bibinfo {pages}
  {371--398}\BibitemShut {NoStop}%
\bibitem [{\citenamefont {Ashton}, \citenamefont {Jarrett},\ and\ \citenamefont
  {Johnson}(2005)}]{ashton_effect_2005}%
  \BibitemOpen
  \bibfield  {author} {\bibinfo {author} {\bibfnamefont {D.~J.}\ \bibnamefont
  {Ashton}}, \bibinfo {author} {\bibfnamefont {T.~C.}\ \bibnamefont {Jarrett}},
  \ and\ \bibinfo {author} {\bibfnamefont {N.~F.}\ \bibnamefont {Johnson}},\
  }\bibfield  {title} {\enquote {\bibinfo {title} {Effect of congestion costs
  on shortest paths through complex networks},}\ }\href@noop {} {\bibfield
  {journal} {\bibinfo  {journal} {Phys. Rev. Lett.}\ }\textbf {\bibinfo
  {volume} {94}},\ \bibinfo {pages} {058701} (\bibinfo {year}
  {2005})}\BibitemShut {NoStop}%
\bibitem [{\citenamefont {Trusina}, \citenamefont {Rosvall},\ and\
  \citenamefont {Sneppen}(2005)}]{trusina_communication_2005}%
  \BibitemOpen
  \bibfield  {author} {\bibinfo {author} {\bibfnamefont {A.}~\bibnamefont
  {Trusina}}, \bibinfo {author} {\bibfnamefont {M.}~\bibnamefont {Rosvall}}, \
  and\ \bibinfo {author} {\bibfnamefont {K.}~\bibnamefont {Sneppen}},\
  }\bibfield  {title} {\enquote {\bibinfo {title} {Communication boundaries in
  networks},}\ }\href@noop {} {\bibfield  {journal} {\bibinfo  {journal} {Phys.
  Rev. Lett.}\ }\textbf {\bibinfo {volume} {94}},\ \bibinfo {pages} {238701}
  (\bibinfo {year} {2005})}\BibitemShut {NoStop}%
\bibitem [{\citenamefont {Estrada}\ and\ \citenamefont
  {Hatano}(2008)}]{estrada_communicability_2008}%
  \BibitemOpen
  \bibfield  {author} {\bibinfo {author} {\bibfnamefont {E.}~\bibnamefont
  {Estrada}}\ and\ \bibinfo {author} {\bibfnamefont {N.}~\bibnamefont
  {Hatano}},\ }\bibfield  {title} {\enquote {\bibinfo {title} {Communicability
  in complex networks},}\ }\href@noop {} {\bibfield  {journal} {\bibinfo
  {journal} {Phys. Rev. E}\ }\textbf {\bibinfo {volume} {77}},\ \bibinfo
  {pages} {036111} (\bibinfo {year} {2008})}\BibitemShut {NoStop}%
\bibitem [{\citenamefont {Huang}\ \emph {et~al.}(2009)\citenamefont {Huang},
  \citenamefont {Chen}, \citenamefont {Lai},\ and\ \citenamefont
  {Pecora}}]{Huang_GenericBehaviour_2009}%
  \BibitemOpen
  \bibfield  {author} {\bibinfo {author} {\bibfnamefont {L.}~\bibnamefont
  {Huang}}, \bibinfo {author} {\bibfnamefont {Q.}~\bibnamefont {Chen}},
  \bibinfo {author} {\bibfnamefont {Y.-C.}\ \bibnamefont {Lai}}, \ and\
  \bibinfo {author} {\bibfnamefont {L.}~\bibnamefont {Pecora}},\ }\bibfield
  {title} {\enquote {\bibinfo {title} {Generic behaviour of master-stability
  functions in coupled nonlinear dynamical sstems},}\ }\href@noop {} {\bibfield
   {journal} {\bibinfo  {journal} {Phys. Rev. E}\ }\textbf {\bibinfo {volume}
  {80}},\ \bibinfo {pages} {036204} (\bibinfo {year} {2009})}\BibitemShut
  {NoStop}%
\bibitem [{\citenamefont {Borgatti}(2005)}]{borgatti_centrality_2005}%
  \BibitemOpen
  \bibfield  {author} {\bibinfo {author} {\bibfnamefont {S.~P.}\ \bibnamefont
  {Borgatti}},\ }\bibfield  {title} {\enquote {\bibinfo {title} {Centrality and
  network flow},}\ }\href@noop {} {\bibfield  {journal} {\bibinfo  {journal}
  {Soc. Networks}\ }\textbf {\bibinfo {volume} {27}},\ \bibinfo {pages}
  {55--71} (\bibinfo {year} {2005})}\BibitemShut {NoStop}%
\bibitem [{\citenamefont {Colizza}\ \emph {et~al.}(2006)\citenamefont
  {Colizza}, \citenamefont {Flammini}, \citenamefont {Serrano},\ and\
  \citenamefont {Vespignani}}]{colizza_detecting_2006}%
  \BibitemOpen
  \bibfield  {author} {\bibinfo {author} {\bibfnamefont {V.}~\bibnamefont
  {Colizza}}, \bibinfo {author} {\bibfnamefont {A.}~\bibnamefont {Flammini}},
  \bibinfo {author} {\bibfnamefont {M.~A.}\ \bibnamefont {Serrano}}, \ and\
  \bibinfo {author} {\bibfnamefont {A.}~\bibnamefont {Vespignani}},\ }\bibfield
   {title} {\enquote {\bibinfo {title} {Detecting rich-club ordering in complex
  networks},}\ }\href@noop {} {\bibfield  {journal} {\bibinfo  {journal} {Nat.
  Phys.}\ }\textbf {\bibinfo {volume} {2}},\ \bibinfo {pages} {110--115}
  (\bibinfo {year} {2006})}\BibitemShut {NoStop}%
\bibitem [{\citenamefont {Go{\~n}i}\ \emph {et~al.}(2014)\citenamefont
  {Go{\~n}i}, \citenamefont {Heuvel}, \citenamefont {Avena-Koenigsberger},
  \citenamefont {Mendizabal}, \citenamefont {Betzel}, \citenamefont {Griffa},
  \citenamefont {Hagmann}, \citenamefont {Corominas-Murtra}, \citenamefont
  {Thiran},\ and\ \citenamefont {Sporns}}]{goni_resting-brain_2014}%
  \BibitemOpen
  \bibfield  {author} {\bibinfo {author} {\bibfnamefont {J.}~\bibnamefont
  {Go{\~n}i}}, \bibinfo {author} {\bibfnamefont {M.~P. v.~d.}\ \bibnamefont
  {Heuvel}}, \bibinfo {author} {\bibfnamefont {A.}~\bibnamefont
  {Avena-Koenigsberger}}, \bibinfo {author} {\bibfnamefont {N.~V.~d.}\
  \bibnamefont {Mendizabal}}, \bibinfo {author} {\bibfnamefont {R.~F.}\
  \bibnamefont {Betzel}}, \bibinfo {author} {\bibfnamefont {A.}~\bibnamefont
  {Griffa}}, \bibinfo {author} {\bibfnamefont {P.}~\bibnamefont {Hagmann}},
  \bibinfo {author} {\bibfnamefont {B.}~\bibnamefont {Corominas-Murtra}},
  \bibinfo {author} {\bibfnamefont {J.-P.}\ \bibnamefont {Thiran}}, \ and\
  \bibinfo {author} {\bibfnamefont {O.}~\bibnamefont {Sporns}},\ }\bibfield
  {title} {\enquote {\bibinfo {title} {Resting-brain functional connectivity
  predicted by analytic measures of network communication},}\ }\href@noop {}
  {\bibfield  {journal} {\bibinfo  {journal} {Proc. Nat. Acad. Sci. USA}\
  }\textbf {\bibinfo {volume} {111}},\ \bibinfo {pages} {833--838} (\bibinfo
  {year} {2014})}\BibitemShut {NoStop}%
\bibitem [{\citenamefont {Avena-Koenigsberger}\ \emph
  {et~al.}(2016)\citenamefont {Avena-Koenigsberger}, \citenamefont
  {Mi\v{s}i\'{c}}, \citenamefont {Hawkins}, \citenamefont {Griffa},
  \citenamefont {andJoaqu\'{i}n Go\~{n}i},\ and\ \citenamefont
  {Sporns}}]{Avena_PathEnsembles_2016}%
  \BibitemOpen
  \bibfield  {author} {\bibinfo {author} {\bibfnamefont {A.}~\bibnamefont
  {Avena-Koenigsberger}}, \bibinfo {author} {\bibfnamefont {B.}~\bibnamefont
  {Mi\v{s}i\'{c}}}, \bibinfo {author} {\bibfnamefont {R.}~\bibnamefont
  {Hawkins}}, \bibinfo {author} {\bibfnamefont {A.}~\bibnamefont {Griffa}},
  \bibinfo {author} {\bibfnamefont {P.~H.}\ \bibnamefont {andJoaqu\'{i}n
  Go\~{n}i}}, \ and\ \bibinfo {author} {\bibfnamefont {O.}~\bibnamefont
  {Sporns}},\ }\bibfield  {title} {\enquote {\bibinfo {title} {Path ensembles
  and a tradeoff between communication efficiency and resilience in the human
  connectome},}\ }\href@noop {} {\bibfield  {journal} {\bibinfo  {journal}
  {Brain Struct. Funct.}\ } (\bibinfo {year} {2016})}\BibitemShut {NoStop}%
\bibitem [{\citenamefont {Harary}\ and\ \citenamefont
  {Schwenk}(1979)}]{harary_spectral_1979}%
  \BibitemOpen
  \bibfield  {author} {\bibinfo {author} {\bibfnamefont {F.}~\bibnamefont
  {Harary}}\ and\ \bibinfo {author} {\bibfnamefont {A.}~\bibnamefont
  {Schwenk}},\ }\bibfield  {title} {\enquote {\bibinfo {title} {The spectral
  approach to determining the number of walks in a graph},}\ }\href@noop {}
  {\bibfield  {journal} {\bibinfo  {journal} {Pacific Journal of Mathematics}\
  }\textbf {\bibinfo {volume} {80}},\ \bibinfo {pages} {443--449} (\bibinfo
  {year} {1979})}\BibitemShut {NoStop}%
\bibitem [{\citenamefont {Bang-Jensen}\ and\ \citenamefont
  {Gutin}(2008)}]{bang-jensen_digraphs_2008}%
  \BibitemOpen
  \bibfield  {author} {\bibinfo {author} {\bibfnamefont {J.}~\bibnamefont
  {Bang-Jensen}}\ and\ \bibinfo {author} {\bibfnamefont {G.~Z.}\ \bibnamefont
  {Gutin}},\ }\href@noop {} {\emph {\bibinfo {title} {Digraphs: {Theory},
  {Algorithms} and {Applications}}}}\ (\bibinfo  {publisher} {Springer},\
  \bibinfo {year} {2008})\BibitemShut {NoStop}%
\bibitem [{\citenamefont {Estrada}, \citenamefont {Hatano},\ and\ \citenamefont
  {Benzi}(2012)}]{estrada_physics_2012}%
  \BibitemOpen
  \bibfield  {author} {\bibinfo {author} {\bibfnamefont {E.}~\bibnamefont
  {Estrada}}, \bibinfo {author} {\bibfnamefont {N.}~\bibnamefont {Hatano}}, \
  and\ \bibinfo {author} {\bibfnamefont {M.}~\bibnamefont {Benzi}},\ }\bibfield
   {title} {\enquote {\bibinfo {title} {The physics of communicability in
  complex networks},}\ }\href@noop {} {\bibfield  {journal} {\bibinfo
  {journal} {Phys. Reps.}\ }\textbf {\bibinfo {volume} {514}},\ \bibinfo
  {pages} {89--119} (\bibinfo {year} {2012})}\BibitemShut {NoStop}%
\bibitem [{\citenamefont {Zamora-L\'opez}\ \emph {et~al.}(2016)\citenamefont
  {Zamora-L\'opez}, \citenamefont {Chen}, \citenamefont {Deco}, \citenamefont
  {Kringelbach},\ and\ \citenamefont {Zhou}}]{zamora_functional_2016}%
  \BibitemOpen
  \bibfield  {author} {\bibinfo {author} {\bibfnamefont {G.}~\bibnamefont
  {Zamora-L\'opez}}, \bibinfo {author} {\bibfnamefont {Y.}~\bibnamefont
  {Chen}}, \bibinfo {author} {\bibfnamefont {G.}~\bibnamefont {Deco}}, \bibinfo
  {author} {\bibfnamefont {M.~L.}\ \bibnamefont {Kringelbach}}, \ and\ \bibinfo
  {author} {\bibfnamefont {C.}~\bibnamefont {Zhou}},\ }\bibfield  {title}
  {\enquote {\bibinfo {title} {Functional complexity emerging from anatomical
  constraints in the brain: the significance of network modularity and
  rich-clubs},}\ }\href {\doibase 10.1038/srep38424} {\bibfield  {journal}
  {\bibinfo  {journal} {Scientific Reports}\ }\textbf {\bibinfo {volume} {6}},\
  \bibinfo {pages} {38424} (\bibinfo {year} {2016})}\BibitemShut {NoStop}%
\bibitem [{\citenamefont {Jones}\ and\ \citenamefont
  {Cercignani}(2010)}]{jones_challenges_2010}%
  \BibitemOpen
  \bibfield  {author} {\bibinfo {author} {\bibfnamefont {D.~K.}\ \bibnamefont
  {Jones}}\ and\ \bibinfo {author} {\bibfnamefont {M.}~\bibnamefont
  {Cercignani}},\ }\bibfield  {title} {\enquote {\bibinfo {title} {Twenty-five
  pitfalls in the analysis of diffusion mri data.}}\ }\href@noop {} {\bibfield
  {journal} {\bibinfo  {journal} {NMR Biomed.}\ }\textbf {\bibinfo {volume}
  {23}},\ \bibinfo {pages} {803 -- 820} (\bibinfo {year} {2010})}\BibitemShut
  {NoStop}%
\bibitem [{\citenamefont {Thomas}\ \emph {et~al.}(2014)\citenamefont {Thomas},
  \citenamefont {Frank}, \citenamefont {Irfanoglu}, \citenamefont {Modi},
  \citenamefont {Saleem}, \citenamefont {Leopold},\ and\ \citenamefont
  {Pierpaoli}}]{thomas_anatomical_2014}%
  \BibitemOpen
  \bibfield  {author} {\bibinfo {author} {\bibfnamefont {C.}~\bibnamefont
  {Thomas}}, \bibinfo {author} {\bibfnamefont {Q.~Y.}\ \bibnamefont {Frank}},
  \bibinfo {author} {\bibfnamefont {M.~O.}\ \bibnamefont {Irfanoglu}}, \bibinfo
  {author} {\bibfnamefont {P.}~\bibnamefont {Modi}}, \bibinfo {author}
  {\bibfnamefont {K.~S.}\ \bibnamefont {Saleem}}, \bibinfo {author}
  {\bibfnamefont {D.~A.}\ \bibnamefont {Leopold}}, \ and\ \bibinfo {author}
  {\bibfnamefont {C.}~\bibnamefont {Pierpaoli}},\ }\bibfield  {title} {\enquote
  {\bibinfo {title} {Anatomical accuracy of brain connections derived from
  diffusion mri tractography is inherently limited},}\ }\href@noop {}
  {\bibfield  {journal} {\bibinfo  {journal} {Proceedings of the National
  Academy of Sciences}\ }\textbf {\bibinfo {volume} {111}},\ \bibinfo {pages}
  {16574--16579} (\bibinfo {year} {2014})}\BibitemShut {NoStop}%
\bibitem [{\citenamefont {Jbabdi}\ \emph {et~al.}(2015)\citenamefont {Jbabdi},
  \citenamefont {Sotiropoulos}, \citenamefont {Haber}, \citenamefont
  {Van~Essen},\ and\ \citenamefont {Behrens}}]{jbabdi_measuring_2015}%
  \BibitemOpen
  \bibfield  {author} {\bibinfo {author} {\bibfnamefont {S.}~\bibnamefont
  {Jbabdi}}, \bibinfo {author} {\bibfnamefont {S.~N.}\ \bibnamefont
  {Sotiropoulos}}, \bibinfo {author} {\bibfnamefont {S.~N.}\ \bibnamefont
  {Haber}}, \bibinfo {author} {\bibfnamefont {D.~C.}\ \bibnamefont
  {Van~Essen}}, \ and\ \bibinfo {author} {\bibfnamefont {T.~E.}\ \bibnamefont
  {Behrens}},\ }\bibfield  {title} {\enquote {\bibinfo {title} {Measuring
  macroscopic brain connections in vivo},}\ }\href@noop {} {\bibfield
  {journal} {\bibinfo  {journal} {Nature neuroscience}\ }\textbf {\bibinfo
  {volume} {18}},\ \bibinfo {pages} {1546--1555} (\bibinfo {year}
  {2015})}\BibitemShut {NoStop}%
\bibitem [{\citenamefont {Freyer}\ \emph {et~al.}(2011)\citenamefont {Freyer},
  \citenamefont {Roberts}, \citenamefont {Becker}, \citenamefont {Robinson},
  \citenamefont {Ritter},\ and\ \citenamefont
  {Breakspear}}]{freyer_biophysical_2011}%
  \BibitemOpen
  \bibfield  {author} {\bibinfo {author} {\bibfnamefont {F.}~\bibnamefont
  {Freyer}}, \bibinfo {author} {\bibfnamefont {J.~A.}\ \bibnamefont {Roberts}},
  \bibinfo {author} {\bibfnamefont {R.}~\bibnamefont {Becker}}, \bibinfo
  {author} {\bibfnamefont {P.~A.}\ \bibnamefont {Robinson}}, \bibinfo {author}
  {\bibfnamefont {P.}~\bibnamefont {Ritter}}, \ and\ \bibinfo {author}
  {\bibfnamefont {M.}~\bibnamefont {Breakspear}},\ }\bibfield  {title}
  {\enquote {\bibinfo {title} {Biophysical mechanisms of multistability in
  resting-state cortical rhythms},}\ }\href@noop {} {\bibfield  {journal}
  {\bibinfo  {journal} {The Journal of Neuroscience}\ }\textbf {\bibinfo
  {volume} {31}},\ \bibinfo {pages} {6353--6361} (\bibinfo {year}
  {2011})}\BibitemShut {NoStop}%
\bibitem [{\citenamefont {Deco}\ and\ \citenamefont
  {Kringelbach}(2016)}]{deco_metastability_2016}%
  \BibitemOpen
  \bibfield  {author} {\bibinfo {author} {\bibfnamefont {G.}~\bibnamefont
  {Deco}}\ and\ \bibinfo {author} {\bibfnamefont {M.~L.}\ \bibnamefont
  {Kringelbach}},\ }\bibfield  {title} {\enquote {\bibinfo {title}
  {Metastability and coherence: extending the communication through coherence
  hypothesis using a whole-brain computational perspective},}\ }\href@noop {}
  {\bibfield  {journal} {\bibinfo  {journal} {Trends in neurosciences}\
  }\textbf {\bibinfo {volume} {39}},\ \bibinfo {pages} {125--135} (\bibinfo
  {year} {2016})}\BibitemShut {NoStop}%
\bibitem [{\citenamefont {Freyer}\ \emph {et~al.}(2012)\citenamefont {Freyer},
  \citenamefont {Roberts}, \citenamefont {Ritter},\ and\ \citenamefont
  {Breakspear}}]{freyer_canonical_2012}%
  \BibitemOpen
  \bibfield  {author} {\bibinfo {author} {\bibfnamefont {F.}~\bibnamefont
  {Freyer}}, \bibinfo {author} {\bibfnamefont {J.~A.}\ \bibnamefont {Roberts}},
  \bibinfo {author} {\bibfnamefont {P.}~\bibnamefont {Ritter}}, \ and\ \bibinfo
  {author} {\bibfnamefont {M.}~\bibnamefont {Breakspear}},\ }\bibfield  {title}
  {\enquote {\bibinfo {title} {A canonical model of multistability and
  scale-invariance in biological systems},}\ }\href@noop {} {\bibfield
  {journal} {\bibinfo  {journal} {PLoS Comput Biol}\ }\textbf {\bibinfo
  {volume} {8}},\ \bibinfo {pages} {e1002634} (\bibinfo {year}
  {2012})}\BibitemShut {NoStop}%
\bibitem [{\citenamefont {Motter}, \citenamefont {Zhou},\ and\ \citenamefont
  {Kurths}(2005)}]{Motter_Enhancing_2005}%
  \BibitemOpen
  \bibfield  {author} {\bibinfo {author} {\bibfnamefont {A.~E.}\ \bibnamefont
  {Motter}}, \bibinfo {author} {\bibfnamefont {C.~S.}\ \bibnamefont {Zhou}}, \
  and\ \bibinfo {author} {\bibfnamefont {J.}~\bibnamefont {Kurths}},\
  }\bibfield  {title} {\enquote {\bibinfo {title} {Enhancing complex-network
  synchronization},}\ }\href@noop {} {\bibfield  {journal} {\bibinfo  {journal}
  {Europhys. Lett.}\ }\textbf {\bibinfo {volume} {69}},\ \bibinfo {pages}
  {334--340} (\bibinfo {year} {2005})}\BibitemShut {NoStop}%
\bibitem [{\citenamefont {di~Bernardo}, \citenamefont {Garofalo},\ and\
  \citenamefont {Sorrentino}(2005)}]{Bernardo_Synch_2005}%
  \BibitemOpen
  \bibfield  {author} {\bibinfo {author} {\bibfnamefont {M.}~\bibnamefont
  {di~Bernardo}}, \bibinfo {author} {\bibfnamefont {F.}~\bibnamefont
  {Garofalo}}, \ and\ \bibinfo {author} {\bibfnamefont {F.}~\bibnamefont
  {Sorrentino}},\ }\bibfield  {title} {\enquote {\bibinfo {title} {Effects of
  degree correlation on the synchronizability of networks of nonlinear
  oscillators},}\ \ }(\bibinfo {organization} {Proc. 44th IEEE Conference on
  Decision and Control, and the European Control Conference WeA14.1},\ \bibinfo
  {year} {2005})\BibitemShut {NoStop}%
\bibitem [{\citenamefont {Atay}, \citenamefont {Biyiko\v{g}lu},\ and\
  \citenamefont {Jost}(2006)}]{Atay_Spectra_2006}%
  \BibitemOpen
  \bibfield  {author} {\bibinfo {author} {\bibfnamefont {F.}~\bibnamefont
  {Atay}}, \bibinfo {author} {\bibfnamefont {T.}~\bibnamefont {Biyiko\v{g}lu}},
  \ and\ \bibinfo {author} {\bibfnamefont {J.}~\bibnamefont {Jost}},\
  }\bibfield  {title} {\enquote {\bibinfo {title} {Network synchronization:
  Spectral versus statistical properties},}\ }\href@noop {} {\bibfield
  {journal} {\bibinfo  {journal} {Physica D}\ }\textbf {\bibinfo {volume}
  {224}},\ \bibinfo {pages} {35--41} (\bibinfo {year} {2006})}\BibitemShut
  {NoStop}%
\bibitem [{\citenamefont {Zhou}\ and\ \citenamefont
  {Kurths}(2006)}]{Zhou_HierarchicalSynch_2006}%
  \BibitemOpen
  \bibfield  {author} {\bibinfo {author} {\bibfnamefont {C.}~\bibnamefont
  {Zhou}}\ and\ \bibinfo {author} {\bibfnamefont {J.}~\bibnamefont {Kurths}},\
  }\bibfield  {title} {\enquote {\bibinfo {title} {Hierarchical synchronization
  in networks of oscillators with heterogeneous degrees},}\ }\href@noop {}
  {\bibfield  {journal} {\bibinfo  {journal} {Chaos}\ }\textbf {\bibinfo
  {volume} {16}},\ \bibinfo {pages} {015104} (\bibinfo {year}
  {2006})}\BibitemShut {NoStop}%
\bibitem [{\citenamefont {Wu}\ \emph {et~al.}(2006)\citenamefont {Wu},
  \citenamefont {Wang}, \citenamefont {Zhou}, \citenamefont {Wang},
  \citenamefont {Zhao},\ and\ \citenamefont {Yang}}]{Wu_SynchroClust_2006}%
  \BibitemOpen
  \bibfield  {author} {\bibinfo {author} {\bibfnamefont {X.}~\bibnamefont
  {Wu}}, \bibinfo {author} {\bibfnamefont {B.}~\bibnamefont {Wang}}, \bibinfo
  {author} {\bibfnamefont {T.}~\bibnamefont {Zhou}}, \bibinfo {author}
  {\bibfnamefont {W.}~\bibnamefont {Wang}}, \bibinfo {author} {\bibfnamefont
  {M.}~\bibnamefont {Zhao}}, \ and\ \bibinfo {author} {\bibfnamefont
  {H.}~\bibnamefont {Yang}},\ }\bibfield  {title} {\enquote {\bibinfo {title}
  {Synchronizability of highly clustered scale-free networks},}\ }\href@noop {}
  {\bibfield  {journal} {\bibinfo  {journal} {Chinese Phys. Lett.}\ }\textbf
  {\bibinfo {volume} {23 (4)}},\ \bibinfo {pages} {1046--1049} (\bibinfo {year}
  {2006})}\BibitemShut {NoStop}%
\bibitem [{\citenamefont {Arenas}\ \emph {et~al.}(2008)\citenamefont {Arenas},
  \citenamefont {D\'iaz-Guilera}, \citenamefont {Kurths}, \citenamefont
  {Moreno},\ and\ \citenamefont {Zhou}}]{Arenas_Review_2008}%
  \BibitemOpen
  \bibfield  {author} {\bibinfo {author} {\bibfnamefont {A.}~\bibnamefont
  {Arenas}}, \bibinfo {author} {\bibfnamefont {A.}~\bibnamefont
  {D\'iaz-Guilera}}, \bibinfo {author} {\bibfnamefont {J.}~\bibnamefont
  {Kurths}}, \bibinfo {author} {\bibfnamefont {Y.}~\bibnamefont {Moreno}}, \
  and\ \bibinfo {author} {\bibfnamefont {C.~S.}\ \bibnamefont {Zhou}},\
  }\bibfield  {title} {\enquote {\bibinfo {title} {Synchronization in complex
  networks},}\ }\href@noop {} {\bibfield  {journal} {\bibinfo  {journal} {Phys.
  Reps.}\ }\textbf {\bibinfo {volume} {469}},\ \bibinfo {pages} {93--153}
  (\bibinfo {year} {2008})}\BibitemShut {NoStop}%
\bibitem [{\citenamefont {Anderson}, \citenamefont {Skare},\ and\ \citenamefont
  {Ashburner}(2003)}]{andersson_howto_2003}%
  \BibitemOpen
  \bibfield  {author} {\bibinfo {author} {\bibfnamefont {J.~L.}\ \bibnamefont
  {Anderson}}, \bibinfo {author} {\bibfnamefont {S.}~\bibnamefont {Skare}}, \
  and\ \bibinfo {author} {\bibfnamefont {J.}~\bibnamefont {Ashburner}},\
  }\bibfield  {title} {\enquote {\bibinfo {title} {How to correct
  susceptibility distortions in spin echo-panar images: application to
  diffusion tensor imaging.}}\ }\href@noop {} {\bibfield  {journal} {\bibinfo
  {journal} {NeuroImage}\ }\textbf {\bibinfo {volume} {20}},\ \bibinfo {pages}
  {870 -- 888} (\bibinfo {year} {2003})}\BibitemShut {NoStop}%
\bibitem [{\citenamefont {Anderson}\ and\ \citenamefont
  {Sotiropoulos}(2016)}]{andersson_integrated_2016}%
  \BibitemOpen
  \bibfield  {author} {\bibinfo {author} {\bibfnamefont {J.}~\bibnamefont
  {Anderson}}\ and\ \bibinfo {author} {\bibfnamefont {S.}~\bibnamefont
  {Sotiropoulos}},\ }\bibfield  {title} {\enquote {\bibinfo {title} {An
  integrated approach to correction for off-resonance effects and subject
  movement in diffusion mr imaging.}}\ }\href@noop {} {\bibfield  {journal}
  {\bibinfo  {journal} {NeuroImage}\ }\textbf {\bibinfo {volume} {125}},\
  \bibinfo {pages} {1063 -- 1078} (\bibinfo {year} {2016})}\BibitemShut
  {NoStop}%
\bibitem [{\citenamefont {Leemans}\ and\ \citenamefont
  {Jones}(2009)}]{leemans_bmatrix_2009}%
  \BibitemOpen
  \bibfield  {author} {\bibinfo {author} {\bibfnamefont {A.}~\bibnamefont
  {Leemans}}\ and\ \bibinfo {author} {\bibfnamefont {D.}~\bibnamefont
  {Jones}},\ }\bibfield  {title} {\enquote {\bibinfo {title} {The b-matrix must
  be rotated when correcting for subject motion in dti data.}}\ }\href@noop {}
  {\bibfield  {journal} {\bibinfo  {journal} {Magn. Reson Med}\ }\textbf
  {\bibinfo {volume} {61}} (\bibinfo {year} {2009})}\BibitemShut {NoStop}%
\bibitem [{\citenamefont {Ersiz}\ \emph {et~al.}(2014)\citenamefont {Ersiz},
  \citenamefont {Arpinar}, \citenamefont {Dreyer},\ and\ \citenamefont
  {Muftuler}}]{Ersoz_quantitative_2014}%
  \BibitemOpen
  \bibfield  {author} {\bibinfo {author} {\bibfnamefont {A.}~\bibnamefont
  {Ersiz}}, \bibinfo {author} {\bibfnamefont {V.~E.}\ \bibnamefont {Arpinar}},
  \bibinfo {author} {\bibfnamefont {S.}~\bibnamefont {Dreyer}}, \ and\ \bibinfo
  {author} {\bibfnamefont {L.~T.}\ \bibnamefont {Muftuler}},\ }\bibfield
  {title} {\enquote {\bibinfo {title} {Quantitative analysis of the efficacy of
  gradient table correction on improving the accuracy of fiber tractography.}}\
  }\href@noop {} {\bibfield  {journal} {\bibinfo  {journal} {Magn. Reson.
  Med.}\ }\textbf {\bibinfo {volume} {72}},\ \bibinfo {pages} {227 -- 236}
  (\bibinfo {year} {2014})}\BibitemShut {NoStop}%
\bibitem [{\citenamefont {Behrens}\ \emph {et~al.}(2003)\citenamefont
  {Behrens}, \citenamefont {Berg}, \citenamefont {Jbabdi}, \citenamefont
  {Rushworth},\ and\ \citenamefont {Woolrich}}]{behrens_characterization_2003}%
  \BibitemOpen
  \bibfield  {author} {\bibinfo {author} {\bibfnamefont {T.~E.}\ \bibnamefont
  {Behrens}}, \bibinfo {author} {\bibfnamefont {H.~J.}\ \bibnamefont {Berg}},
  \bibinfo {author} {\bibfnamefont {S.}~\bibnamefont {Jbabdi}}, \bibinfo
  {author} {\bibfnamefont {M.~F.~S.}\ \bibnamefont {Rushworth}}, \ and\
  \bibinfo {author} {\bibfnamefont {M.~W.}\ \bibnamefont {Woolrich}},\
  }\bibfield  {title} {\enquote {\bibinfo {title} {Charaterization and
  propagation of uncertainty in diffusion weighted mr imaging.}}\ }\href@noop
  {} {\bibfield  {journal} {\bibinfo  {journal} {Magn. Reson. Med.}\ }\textbf
  {\bibinfo {volume} {72}},\ \bibinfo {pages} {227 -- 236} (\bibinfo {year}
  {2003})}\BibitemShut {NoStop}%
\bibitem [{\citenamefont {Tzourio-Mazoyera}\ \emph {et~al.}(0215)\citenamefont
  {Tzourio-Mazoyera}, \citenamefont {Landeaub}, \citenamefont
  {Papathanassioua}, \citenamefont {Crivelloa}, \citenamefont {Etarda},
  \citenamefont {Delcroixa}, \citenamefont {Mazoyerc},\ and\ \citenamefont
  {Joliot}}]{tzourio_AALparcellation_2002}%
  \BibitemOpen
  \bibfield  {author} {\bibinfo {author} {\bibfnamefont {N.}~\bibnamefont
  {Tzourio-Mazoyera}}, \bibinfo {author} {\bibfnamefont {B.}~\bibnamefont
  {Landeaub}}, \bibinfo {author} {\bibfnamefont {D.}~\bibnamefont
  {Papathanassioua}}, \bibinfo {author} {\bibfnamefont {F.}~\bibnamefont
  {Crivelloa}}, \bibinfo {author} {\bibfnamefont {O.}~\bibnamefont {Etarda}},
  \bibinfo {author} {\bibfnamefont {N.}~\bibnamefont {Delcroixa}}, \bibinfo
  {author} {\bibfnamefont {B.}~\bibnamefont {Mazoyerc}}, \ and\ \bibinfo
  {author} {\bibfnamefont {M.}~\bibnamefont {Joliot}},\ }\bibfield  {title}
  {\enquote {\bibinfo {title} {Automated anatomical labeling of activations in
  {SPM} using a macroscopic anatomical parcellation of the {MNI} {MRI}
  single-subject brain},}\ }\href@noop {} {\bibfield  {journal} {\bibinfo
  {journal} {Neuroimage}\ }\textbf {\bibinfo {volume} {1}} (\bibinfo {year}
  {200215})}\BibitemShut {NoStop}%
\bibitem [{\citenamefont {Smith}\ \emph {et~al.}(2001)\citenamefont {Smith},
  \citenamefont {{De Stefano}}, \citenamefont {Jenkinson},\ and\ \citenamefont
  {Matthews}}]{smith_normalized_2001}%
  \BibitemOpen
  \bibfield  {author} {\bibinfo {author} {\bibfnamefont {S.~M.}\ \bibnamefont
  {Smith}}, \bibinfo {author} {\bibfnamefont {N.}~\bibnamefont {{De Stefano}}},
  \bibinfo {author} {\bibfnamefont {M.}~\bibnamefont {Jenkinson}}, \ and\
  \bibinfo {author} {\bibfnamefont {P.~M.}\ \bibnamefont {Matthews}},\
  }\bibfield  {title} {\enquote {\bibinfo {title} {Normalized accurate
  measurement of longitudinal brain change.}}\ }\href@noop {} {\bibfield
  {journal} {\bibinfo  {journal} {J. Comput. Assist. Tomogr.}\ }\textbf
  {\bibinfo {volume} {25}},\ \bibinfo {pages} {466 -- 475} (\bibinfo {year}
  {2001})}\BibitemShut {NoStop}%
\bibitem [{\citenamefont {Behrens}\ \emph {et~al.}(2007)\citenamefont
  {Behrens}, \citenamefont {Berg}, \citenamefont {Jbabdi}, \citenamefont
  {Rushworth},\ and\ \citenamefont {Woolrich}}]{behrens_probDTI_2007}%
  \BibitemOpen
  \bibfield  {author} {\bibinfo {author} {\bibfnamefont {T.}~\bibnamefont
  {Behrens}}, \bibinfo {author} {\bibfnamefont {H.}~\bibnamefont {Berg}},
  \bibinfo {author} {\bibfnamefont {S.}~\bibnamefont {Jbabdi}}, \bibinfo
  {author} {\bibfnamefont {M.}~\bibnamefont {Rushworth}}, \ and\ \bibinfo
  {author} {\bibfnamefont {M.}~\bibnamefont {Woolrich}},\ }\bibfield  {title}
  {\enquote {\bibinfo {title} {Probabilistic diffusion tractography with
  multiple fibre orientations: What can we gain?}}\ }\href@noop {} {\bibfield
  {journal} {\bibinfo  {journal} {Neuroimage}\ }\textbf {\bibinfo {volume}
  {34}} (\bibinfo {year} {2007})}\BibitemShut {NoStop}%
\bibitem [{\citenamefont {{de Reus}}\ and\ \citenamefont {{van den
  Heuvel}}(2013)}]{reus_falsepositives_2013}%
  \BibitemOpen
  \bibfield  {author} {\bibinfo {author} {\bibfnamefont {M.~A.}\ \bibnamefont
  {{de Reus}}}\ and\ \bibinfo {author} {\bibfnamefont {M.}~\bibnamefont {{van
  den Heuvel}}},\ }\bibfield  {title} {\enquote {\bibinfo {title} {Estimating
  false positives and negatives in brain networks},}\ }\href@noop {} {\bibfield
   {journal} {\bibinfo  {journal} {NeuroImage}\ }\textbf {\bibinfo {volume}
  {70}},\ \bibinfo {pages} {402 -- 409} (\bibinfo {year} {2013})}\BibitemShut
  {NoStop}%
\bibitem [{\citenamefont {Patel}\ \emph {et~al.}(2014)\citenamefont {Patel},
  \citenamefont {Kundu}, \citenamefont {Rubinov}, \citenamefont {Jones},
  \citenamefont {V\'erts}, \citenamefont {Ersche}, \citenamefont {Suckling},\
  and\ \citenamefont {Bullmore}}]{patel_wavelet_2014}%
  \BibitemOpen
  \bibfield  {author} {\bibinfo {author} {\bibfnamefont {A.~X.}\ \bibnamefont
  {Patel}}, \bibinfo {author} {\bibfnamefont {P.}~\bibnamefont {Kundu}},
  \bibinfo {author} {\bibfnamefont {M.}~\bibnamefont {Rubinov}}, \bibinfo
  {author} {\bibfnamefont {P.~S.}\ \bibnamefont {Jones}}, \bibinfo {author}
  {\bibfnamefont {P.~E.}\ \bibnamefont {V\'erts}}, \bibinfo {author}
  {\bibfnamefont {K.~D.}\ \bibnamefont {Ersche}}, \bibinfo {author}
  {\bibfnamefont {J.}~\bibnamefont {Suckling}}, \ and\ \bibinfo {author}
  {\bibfnamefont {E.~T.}\ \bibnamefont {Bullmore}},\ }\bibfield  {title}
  {\enquote {\bibinfo {title} {A wavelet method for modeling and despiking
  motion artifacts from resting-state fmri time series.}}\ }\href@noop {}
  {\bibfield  {journal} {\bibinfo  {journal} {NeuroImage}\ }\textbf {\bibinfo
  {volume} {95}},\ \bibinfo {pages} {287 -- 304} (\bibinfo {year}
  {2014})}\BibitemShut {NoStop}%
\bibitem [{\citenamefont {Glerean}\ \emph {et~al.}(2012)\citenamefont
  {Glerean}, \citenamefont {Salmi}, \citenamefont {Lahnakoski}, \citenamefont
  {J{\"a}{\"a}skel{\"a}inen},\ and\ \citenamefont
  {Sams}}]{glerean_functional_2012}%
  \BibitemOpen
  \bibfield  {author} {\bibinfo {author} {\bibfnamefont {E.}~\bibnamefont
  {Glerean}}, \bibinfo {author} {\bibfnamefont {J.}~\bibnamefont {Salmi}},
  \bibinfo {author} {\bibfnamefont {J.~M.}\ \bibnamefont {Lahnakoski}},
  \bibinfo {author} {\bibfnamefont {I.~P.}\ \bibnamefont
  {J{\"a}{\"a}skel{\"a}inen}}, \ and\ \bibinfo {author} {\bibfnamefont
  {M.}~\bibnamefont {Sams}},\ }\bibfield  {title} {\enquote {\bibinfo {title}
  {Functional magnetic resonance imaging phase synchronization as a measure of
  dynamic functional connectivity},}\ }\href@noop {} {\bibfield  {journal}
  {\bibinfo  {journal} {Brain connectivity}\ }\textbf {\bibinfo {volume} {2}},\
  \bibinfo {pages} {91--101} (\bibinfo {year} {2012})}\BibitemShut {NoStop}%
\end{thebibliography}
\end{document}